\documentclass[12pt]{article}
\pdfoutput=1
\usepackage{putex}
\usepackage[normalem]{ulem}
\usepackage{array}
\newcolumntype{L}[1]{>{\raggedright\let\newline\\\arraybackslash\hspace{0pt}}m{#1}}
\newcolumntype{C}[1]{>{\centering\let\newline\\\arraybackslash\hspace{0pt}}m{#1}}
\newcolumntype{R}[1]{>{\raggedleft\let\newline\\\arraybackslash\hspace{0pt}}m{#1}}

\makeatletter
\renewcommand*\env@matrix[1][*\c@MaxMatrixCols c]{%
  \hskip -\arraycolsep
  \let\@ifnextchar\new@ifnextchar
  \array{#1}}
\makeatother

\newcommand{\dds}{\,{\mathfrak D}}

\begin{document}

\preprint{PUPT-2492}

\institution{PU}{Joseph Henry Laboratories, Princeton University, Princeton, NJ 08544, USA}
\institution{PCTS}{Princeton Center for Theoretical Science, Princeton University, Princeton, NJ 08544, USA}
\institution{IPMU}{Kavli IPMU (WPI), UTIAS, The University of Tokyo, Kashiwa, Chiba 277-8583, Japan}

\title{Monopole operators from the $4-\epsilon$ expansion}

\authors{Shai M.~Chester,\worksat{\PU} M\'ark Mezei,\worksat{\PCTS}  Silviu S.~Pufu,\worksat{\PU} and Itamar Yaakov\worksat{\PU,\IPMU} }

\abstract{
Three-dimensional quantum electrodynamics with $N$ charged fermions contains monopole operators that have been studied perturbatively at large $N$.  
Here, we initiate the study of these monopole operators in the $4-\epsilon$ expansion  by generalizing them to codimension-3 defect operators in $d = 4-\epsilon$ spacetime dimensions.  Assuming the infrared dynamics is described by an interacting CFT, we define the ``conformal weight'' of these operators in terms of the free energy density on $S^2 \times \HH^{2-\epsilon}$ in the presence of magnetic flux through the $S^2$, and calculate this quantity to next-to-leading order in $\epsilon$.  Extrapolating the conformal weight to $\epsilon = 1$ gives an estimate of the scaling dimension of the monopole operators in $d=3$ that does not rely on the $1/N$ expansion. We also perform the computation of the conformal weight in the large $N$ expansion for any $d$ and find agreement between the large $N$ and the small $\epsilon$ expansions in their overlapping regime of validity.  
}

\date{November 2015}

\maketitle

\tableofcontents

\section{Introduction and Summary}

\subsection{Motivation and setup}
A fascinating class of relatively simple, yet sufficiently non-trivial, quantum field theories (QFTs) in $2+1$ space-time dimensions can be obtained by coupling a $U(1)$ gauge theory to charged matter. These theories share certain features with 3+1 dimensional QCD, such as asymptotic freedom and flowing to strong coupling in the infrared, but are easier to study because their gauge group is abelian. In addition to the local gauge-invariant operators that can be constructed as polynomials in the fundamental fields appearing in the Lagrangian and their derivatives, such QFTs contain also monopole operators \cite{Polyakov:1975rs}. These are local, gauge invariant operators distinguished by the fact that they carry non-zero charge under a topological global symmetry group $U(1)_\text{top}$, whose conserved current is 
 \es{ConsCurrent}{
   j^\mu = \frac{1}{8 \pi} \epsilon^{\mu\nu\rho} F_{\nu \rho} \,.
 }
Here, $F_{\nu \rho} = \partial_\nu A_\rho - \partial_\rho A_\nu$ is the field strength of the Abelian gauge field $A_\mu$.   The current \eqref{ConsCurrent} is conserved due to the Bianchi identity satisfied by the gauge field strength. The existence of operators charged under $U(1)_\text{top}$ is intimately tied to the non-trivial topology of the gauge group, in particular to its non-vanishing fundamental group $\pi_1(U(1)) \cong \Z$. The Dirac quantization condition implies that the conserved charge 
 \es{ConsCharge}{
   q = \int d^2 x \, j_0 
 }
satisfies $q \in \Z/2$, or equivalently, that the integrated magnetic flux through a small two sphere surrounding a monopole operator is $4 \pi q \in 2 \pi \Z$. Monopole operators can play an important role in the dynamics of these theories---see, for instance,~\cite{Wen:1993zza, Chen:1993cd, Sachdev97, Rantner01, Rantner:2002zz, Motrunich:2003fz, SVBSF, SBSVF, Hermele, Hermele05, Ran06, Kaul08, Kaul:2008xw, Sachdev:2010uz} for examples arising in various condensed matter systems.

Monopole operators are remarkably difficult to study.  Even when the QFT of interest has a weakly coupled description, which is the case when the number of charged matter fields is very large, monopole operators cannot be studied using traditional techniques.  Progress can be made in the deep infrared (IR), if one assumes that the IR limit of the renormalization group (RG) flow is described by a conformal field theory (CFT) with unbroken $U(1)_\text{top}$ symmetry. One can then use the state-operator correspondence to map a monopole operator of topological charge $q$ inserted at $\vec{x} = 0$ to the ground state on $S^2\times \R$ in the presence of magnetic flux equal to $4 \pi q$ though the $S^2$ \cite{Borokhov:2002ib}.  The scaling dimension of the monopole operator is determined by the ground state energy (or free energy) on $S^2\times \R$.\footnote{  The scaling dimensions of monopole operators were first computed in~\cite{Murthy:1989ps} in the $\CP^{N_b-1}$ model without making use of the state-operator correspondence.  The computation in \cite{Murthy:1989ps} was only at leading order in $1/N_b$.}  At weak coupling, this ground state energy can be computed by performing a saddle point approximation around the configuration where the magnetic flux is uniformly distributed on the $S^2$.

Monopole operators in conformal field theories have been studied using the $1/N$ expansion \cite{Borokhov:2002ib,Pufu:2013vpa,Dyer:2013fja}. BPS monopole operators, which can exist in theories with $\mathcal{N}\ge 2$ supersymmetry, have been studied using supersymmetric localization \cite{Benini:2009qs,Benini:2011cma}. In this paper, we study monopole operators using the $4-\epsilon$ expansion.\footnote{It is also interesting to approach the problem of monopole scaling dimensions using numerical methods such as Quantum Monte Carlo~\cite{2013PhRvL.111m7202B, 2015arXiv150205128K} and the conformal bootstrap \cite{Rattazzi:2008pe,Chester:2015}.
}\footnote{Besides the $1/N$ and $\epsilon$ expansions, monopole operators can be studied in the recently proposed large charge expansion~\cite{Hellerman:2015nra}, where it was observed that a subleading term in the large charge limit of operator dimensions can be calculated using effective field theory considerations.  It would be very interesting to understand if it is possible to calculate the coefficients of the other terms in the large charge expansion using this method.}  In this approach, one formally continues the CFT to $d=4-\epsilon$ dimensions and takes $\epsilon$ to be small.
The $4-\epsilon$ expansion is a well-established method for studying more conventional operators \cite{Wilson:1973jj}. It has recently been used to provide an approximation for the $S^3$ free energy of various $3d$ CFTs with vector-like matter \cite{Giombi:2015haa,Fei:2015oha,Fei:2014yja,Giombi:2014xxa,Fei:2014xta,Fei:2015kta}. To the best of our knowledge, the only other example of a disorder operator that has been studied in the  $4-\epsilon$ expansion is the twist defect of the $3d$ Ising model~\cite{Gaiotto:2013nva}.

The generalization of monopole operators to $d=4-\epsilon$ dimensions requires a little thought.  In going away from $d=3$, it is natural to retain the notion of a conserved $U(1)_\text{top}$ quantum number and to think of a charge $q$ monopole operator as a codimension-$3$ defect operator that creates magnetic flux $4 \pi q$ through the $S^2$ that surrounds the codimension-$3$ defect. In $d=4$ the conserved current in \eqref{ConsCurrent} becomes a conserved $2$-form equal to the Hodge dual of the gauge field strength---see \cite{Gaiotto:2014kfa} for a description of such higher-form symmetries. The generalized monopole operator is a 't Hooft line operator that carries non-vanishing charge under a global symmetry associated to a conserved current two-form.

Having defined these codimension-3 operators in $d = 4-\epsilon$ dimensions, we would also like to define a ``conformal weight'' that characterizes the transformation properties of these operators under dilatations and agrees with the scaling dimension of the monopole operators when $\epsilon=1$.  As in~\cite{Kapustin:2005py}, we consider a planar defect operator extending along a flat $\R^{1-\epsilon}$ that passes through the origin.  Using cylindrical coordinates around it, we can make a conformal mapping
\es{ConfMap}
{
ds^2=ds^2_{\R^{1-\epsilon}}+ dr^2+r^2 ds_{S^2}^2 \quad \to \quad d\tilde s^2={ds^2_{\R^{1-\epsilon}}+ dr^2\ov r^2}+ ds_{S^2}^2
}
from $\R^{4-\epsilon}$ to $S^2\times \HH^{2 - \epsilon}$.  Note that the defect operator sitting at $r=0$ is mapped to the conformal boundary of $\HH^{2-\epsilon}$.  This configuration has magnetic flux $4 \pi q$ through $S^2$, and some free energy  density $F^{(q)}/ \Vol(\HH^{2-\epsilon})$ in the presence of this flux. We normalize this quantity by considering the difference

 \es{DeltaFDef}{
   \frac{\Delta F^{(q)} }{ \Vol(\HH^{2 -\epsilon})} \equiv  \frac{ F^{(q)} - F^{(0)} }{ \Vol(\HH^{2-\epsilon}) } 
 }
in the free energy density with the curvature radii set to $1$ between the configuration with magnetic flux $4 \pi q$  threading $S^2$ and the vacuum. In four dimensions, this difference gives the scaling weight $h_q$ of the 't Hooft line operator~\cite{Kapustin:2005py}, while in three dimensions it gives the scaling dimension $\Delta_q$ of the monopole operator.\footnote{Note that in three dimensions~\eqref{ConfMap} describes $S^2\times \R$, hence we recover the conventional state-operator correspondence.}

In the rest of this paper, we restrict our attention to quantum electrodynamics with $N$ flavors of two-component complex fermions of unit charge under the $U(1)$ gauge group.\footnote{The theory of $N_b$ complex scalars coupled to a $U(1)$ gauge theory only has a non-trivial fixed point in the $4-\epsilon$ expansion for $N_b\geq 183$~\cite{Moshe:2003xn}, which makes the $4-\epsilon$ expansion unsuitable for studying such theories at small $N_b$. \label{footnote4}}  We choose $N$ to be even in order to avoid a parity anomaly in $d=3$ \cite{Redlich:1983dv, Redlich:1983kn,Niemi:1983rq}. The bare Lagrangian of this theory on a $d$-dimensional curved manifold is
\es{firstActionIntro}{
\mathcal{S}=&\int d^{d}x\ \sqrt{g}\left[\frac{1}{4e_0^2}F_{\mu\nu} F^{\mu\nu}  + \sum_{\alpha = 1}^{N/2} \bar\psi_\alpha\left(i\slashed{\nabla}+\slashed{A}\right)\psi_\alpha\right] \,,
}
where $e_0$ is the bare gauge coupling, and we grouped our $N$ two-component fermions into $N/2$ four-component fermions. 

\subsection{Summary of results}

Before delving into the details of our computation, let us summarize our results. 

\subsubsection{Results from the $4-\epsilon$ expansion}

The result of the $4-\epsilon$ expansion is reached in two steps. The first step is to expand the free energy at small $e_0^2$. The expansion takes the form:
\es{EpsilonExp0}{
{F^{(q)}\ov \Vol\le(\HH^{2-\epsilon}\ri) }={f_\text{classical}^{(q)}\ov e_0^2}+f_\text{1-loop}^{(q)}+e_0^2\,  f_\text{2-loops}^{(q)}+ \dots\,,
}
where, as we will explain, these terms represent contributions from the classical action, the one-loop determinant of the fermions, the two-loop current-current vacuum diagram, etc.   All these terms are explicit functions of $\epsilon$, and we obtain them by first expanding the relevant diagrams in $\epsilon$ and then using zeta-function regularization to regularize any divergences.  We only obtain these terms up to some small order in the small $\epsilon$ expansion.  These terms also depend on the ultraviolet (UV) cutoff scale $\mu$ that appears implicitly in the definition of the theory and on the curvature radii of $S^2$ and $\HH^{2-\epsilon}$, which are both taken to be equal to $R$. The dependence on $\mu$ and $R$ is only through the dimensionless combination $\mu R$. 

The second step in our calculation is to find the value of $e_0$ corresponding to an RG fixed point and plug it in \eqref{EpsilonExp0}.  This critical value is customarily found from the vanishing of the  beta function.  However, beta functions are often scheme dependent, and one would have to make sure that the scheme used for obtaining the beta function matches the regularization procedure used in other parts of the computation.    To circumvent this potential annoyance, we find the critical value of the bare coupling $e_0$ by requiring that $\Delta F^{(q)}$ is $\mu$-independent.  From this requirement, we obtain:
\es{RenormalizedCouplingIntro}{
e_0&=\le({\mu^2\ov 4 \pi e^{-\gamma} }\ri)^{\epsilon/4}\, \sqrt{12 \pi^2 \epsilon \ov N}\le(1-{9\ov 8 N}\epsilon+\dots\ri)\,,
}
where $\gamma$ is the Euler-Mascheroni constant.  By plugging this into~\eqref{EpsilonExp0} we turn the double expansion in $e_0$ and $\epsilon$ into an expansion in only $\epsilon$:
\es{EpsilonExp}{
{\Delta F^{(q)}\ov \Vol\le(\HH^{2-\epsilon}\ri) }={f_0^{(q)}\ov \epsilon}+f_1^{(q)}+ f_2^{(q)} \epsilon + \dots\,.
}
We find that the first two coefficients are
\es{fiq}{
 f_{0}^{(q)}={N\, q^2\ov 6 \pi} \,, \quad\quad f_{1}^{(q)}=N {\mathfrak f}(q)+{3\, q^2\ov 8 \pi} \,,
 }
and ${\mathfrak f}(q)$ is given in Table~\ref{aTable} for several values of $q$ (a general formula is given by \eqref{fExplicitly} in Appendix~\ref{explicit}). In obtaining \eqref{RenormalizedCouplingIntro}, we needed to determine the divergent piece of $f_\text{2-loop}^{(q)}$ from~\eqref{EpsilonExp0}.  It would be interesting to also determine the finite part of $f_{\text{2-loop}}^{(q)}$, which would be needed to compute the next coefficient, $f_2^{(q)}$ in~\eqref{EpsilonExp}, but we leave this computation for future work. 
 \begin{table}[!h]
\begin{center} \begin{tabular}{c|c}
 $q$ & ${\mathfrak f}(q)$ \\
 \hline
 $1/2$ & $0.06366$\\

$1$& $0.17717$\\

$3/2$&$0.32968$\\

$2$&$0.51509$\\

$5/2$&$0.72917$\\

$\vdots$ & $\vdots$ \\

$10$&$6.00321$ \\

$\vdots$ & $\vdots$ 

  \end{tabular}
  \end{center}
  \caption{The function ${\mathfrak f}(q)$ defined in~\eqref{fiq} for several values of $q$.\label{aTable}}
 \end{table}

Our primary interest in studying the codimension-3 defect operators in $4-\epsilon$ dimensions is to provide a setup for another approximation scheme for the scaling dimension of monopole operators in three dimensions that does not rely on large $N$.  Such an approximation can be performed by extrapolating \eqref{EpsilonExp0} to $\epsilon=1$, using, for instance, a Pad\'e resummation. Since we have computed only the first two terms in \eqref{EpsilonExp0}, we cannot yet perform a meaningful resummation, but we hope that such an extrapolation can eventually be done when more terms in \eqref{EpsilonExp0} become available.

Our setup contains two parameters that can be used to gain some insight into the resummation of the $4-\epsilon$ expansion, namely the number of matter fields $N$ and the monopole charge $q$, which we now discuss.

\subsubsection{Large $N$ limit}

The free energy on  $S^2\times \HH^{2-\epsilon}$ can be evaluated in the large $N$ expansion for any fixed $d=4-\epsilon$, with $\epsilon$ not necessarily small.  In $d=3$, this approach was used for theories with fermionic matter in~\cite{Pufu:2013vpa,Dyer:2013fja}.  The structure of the expansion is:
\es{NExp}{
{\Delta F_{}^{(q)}\ov \Vol\le(\HH^{2-\epsilon}\ri) }=N_{} \, g_{0}^{(q)}(\epsilon)+ g_{1}^{(q)}(\epsilon)+ \frac{g_{2}^{(q)}(\epsilon) }{N} + \dots\,.
}

From the consistency of~\eqref{NExp} with~\eqref{EpsilonExp} in the overlapping region of large $N$ and small $\epsilon$ we expect that
\es{Consistency}{
g_{0}^{(q)}(\epsilon)&={ q^2\ov 6 \pi}\,{1\ov \epsilon}+ {\mathfrak f}(q)+O(\epsilon) \,, \\
g_{1}^{(q)}(\epsilon)&={3\, q^2\ov 8 \pi}+ O(\epsilon) \,.
}
In Section~\ref{largeN}, we compute explicitly $g_{0}^{(q)}(\epsilon)$, and by expanding it at small $\epsilon$ we do indeed recover \eqref{Consistency}.  In other words, the $N \to \infty$ and $\epsilon \to 0$ limits commute.  The fact that these limits commute may seem rather trivial, but the derivation we present uncovers an interesting subtlety.   Indeed, if $\epsilon > 0$, the Maxwell term in the action \eqref{firstActionIntro} is irrelevant and has to be dropped, hence the entire contribution to $g_{0}^{(q)}(\epsilon)$ comes from the regularized $1$-loop determinant of the fermions.  This is very different from the $4-\epsilon$ expansion at finite $N$ \eqref{EpsilonExp}, where the Maxwell term cannot be ignored and contributes the term $\propto 1/\epsilon$ at small~$\epsilon$.  (Said differently, the $\epsilon \to 0$ limit and the regularization of the infinities do not commute.)  Nevertheless, expanding the regularized fermion determinant at small $\epsilon$, we do reproduce the classical contribution of the Maxwell term.  We find this agreement quite remarkable! 

Since the large $N$ and small $\epsilon$ expansions commute, it makes sense to consider the double expansion in $1/N$ and $\epsilon$.
 \begin{table}[htp]
\begin{center}
\begin{tabular}{C{1cm}C{1cm}C{1cm}C{1cm}C{1cm}C{1cm}}
 $\dfrac{N}{\epsilon}$ & $N$ & $N \epsilon$ & $N \epsilon^2$ & $N\epsilon^3$ & $\ldots$ \\[7pt]
  & $1$ & $\epsilon$ & $\epsilon^2$ & $\epsilon^3$ & $\ldots$ \\[20pt]
  & & $\dfrac{\epsilon}{N}$ & $\dfrac{\epsilon^2}{N}$ & $\dfrac{\epsilon^3}{N}$ & $\ldots$ \\[20pt]
  & & & $\dfrac{\epsilon^2}{N^2}$ & $\dfrac{\epsilon^3}{N^2}$ & $\ldots$ \\[20pt]
  & & & & $\dfrac{\epsilon^3}{N^3}$ & $\ldots$ \\[20pt]
  & & & & $\vdots$ & $\ddots$
\end{tabular}
\end{center}
\caption{The dependence on $N$ and $\epsilon$ of the non-zero terms in the double expansion in $1/N$ and $\epsilon$.}
\label{Double}
\end{table}%
The non-zero terms in this expansion are shown in Table~\ref{Double}.  As can be seen from this table, at fixed order $\epsilon^k$ the $1/N$ expansion is finite;  it starts with a term of order $N$ and ends with a term of order $1/N^k$.  (This pattern is already visible in the expansions given above in \eqref{EpsilonExp} and \eqref{NExp}.)  The fact that at fixed $\epsilon$, the $1/N$ expansion terminates is a property of the fermionic theory---such a feature would not arise in similar computations in theories with bosonic matter.

The difference between the large $N$ and small $\epsilon$ expansions is in how precisely we resum the terms in Table~\ref{Double}.  In the large $N$ expansion at fixed $\epsilon$, one essentially resums each row of Table~\ref{Double} first and then extrapolates the sum of the first few rows to finite $N$.  (See \cite{Pufu:2013vpa,Dyer:2013fja} for results in $d=3$ from the large $N$ expansion.) In the small $\epsilon$ expansion at fixed $N$, one resums each column of Table~\ref{Double} first and then extrapolates the sum of these columns to finite~$\epsilon$.  It is of course possible to contemplate other ways of resumming the terms in Table~\ref{Double}---for instance, first resumming each NE-SW diagonal or, alternatively, first resumming each NW-SE diagonal.  Since the full resummation is believed to be only an asymptotic series, it could happen, in principle, that some ways of resumming the entries of Table~\ref{Double} would provide better approximations than others.\footnote{It is worth noting that in \cite{Giombi:2015haa} it was noticed that in the same theory as the one considered here, the $\epsilon$ expansion and the large $N$ expansion resummations of the free energy on $S^{4-\epsilon}$ give similar results.  In the double expansion in $\epsilon$ and $1/N$, the $S^{4-\epsilon}$ free energy has a similar structure as the free energy on $S^2 \times \HH^{2 - \epsilon}$ in the presence of monopole flux:  in particular, for fixed order in $\epsilon$, the $1/N$ expansion contains only a finite number of terms.  So based on this evidence, one may conjecture that the resummations of the free energy on $S^2 \times \HH^{2 - \epsilon}$ using the $\epsilon$ expansion and the $1/N$ expansion would give similar results.}

\subsubsection{Large $q$ limit}

Our setup has another parameter, $q$, which takes only discrete values but which can be taken to be large. In the limit $q\to\infty$ we can derive some analytic formulas. In~\cite{Dyer:2015zha} it was pointed out that $q \to \infty$  is a flat space limit: it corresponds to a uniform large magnetic field on $S^2$, which makes particles move on highly localized Landau levels that are ignorant about the curvature of $S^2$.\footnote{The particles are free to move in the rest of the $1-\epsilon$ spatial directions, but this doesn't change our conclusions.} Then from flat space dimensional analysis we conclude that
\es{Largeq}{
{\Delta F^{(q)}\ov \Vol\le(\HH^{2-\epsilon}\ri) }\propto q^{(4-\epsilon)/2}+\dots\,.
}
Ref.~\cite{Hellerman:2015nra} gave an effective field theory argument to the same effect in $d=3$. 

Let us see how our results match these expectations. Although we only know ${\mathfrak f}(q)$ in terms of a sum and integral, we can derive its large $q$ behavior, from which we get:
\es{Largeq3}{
\frac{\Delta F^{(q)}}{\text{Vol}(\HH^{2-\epsilon})} 
=& {\cal N}(\epsilon) \, q^2 \le[1-{\epsilon\ov 2} \log\left(q\right)+O\le({1\ov q},\epsilon^2\ri)\ri] \,, \\
 {\cal N}(\epsilon)\equiv& {N \ov 6\pi\epsilon}\le[1+ \le(\frac12 \log\left(2 \pi e^{-\gamma}\, A^{12}\right)+{9\ov 4N}\ri)\epsilon+O(\epsilon^2)\ri] \,,
}
where $\gamma$ is the Euler-Mascheroni constant and $A$ is the Glaisher constant. From this expression we readily see that $q^2\le(1-{\epsilon\ov 2}\log q+\dots\ri)$ resums to~\eqref{Largeq}, and we obtain an expansion for the $\epsilon$-dependent prefactor. (Of course, this analysis does not say anything about what the correct resummation for the prefactor might be.) The large-$q$ expansion thus provides another clue as to how to resum the $4-\epsilon$ expansion.  

It is worth mentioning that the first few term of the large $q$ expansion written down in~\eqref{largqFinalFerm} approximates the ${\mathfrak f}(q)$ values in Table~\ref{aTable} to 4 digits precision even at $q = 1/2$, so in practice~\eqref{largqFinalFerm} is valid for all allowed  values of $q$. We do not currently have an explanation for this observation.

\subsection{Organization of the paper}
The rest of this paper is organized as follows.  Section~\ref{EXPANSION} represents the main part of this paper where we compute the free energy density in the presence of a monopole background on $S^2 \times \HH^{2 - \epsilon}$.  In Section~\ref{largeN} we perform the leading large $N$ expansion computation for arbitrary $\epsilon$ and compare it to the computation of Section~\ref{EXPANSION}.  In Section~\ref{sec:global} we remark on how the defect operators transform under the global symmetries of the theory. Several technical details of our computation are included in the Appendices.

\section{Monopole operators in the $\epsilon$ expansion}
\label{EXPANSION}

\subsection{Conventions}\label{setup}

We will investigate $U(1)$ gauge theories with $N/2$ four-component Dirac fermions of unit charge. In $d=4$, these theories have an $SU(N/2)_L\times SU(N/2)_R$ flavor symmetry. In $d=3$, the symmetry is enhanced to $SU(N)$. The defect operators are expected to transform in representations of this symmetry.\footnote{ We discuss global symmetries more extensively in Section~\ref{sec:global}. }  The action on a manifold of dimension $d=4-\epsilon$  is
\es{firstAction}{
\mathcal{S}=&\int d^{4-\epsilon}x\ \sqrt{g}\left[\frac{1}{4e_0^2}(\mathcal{F}_{\mu\nu}+f_{\mu\nu})(\mathcal{F}^{\mu\nu}+f^{\mu\nu}) +\sum_{\alpha=1}^{N/2} \bar\psi_\alpha\left(i\slashed{\nabla}+\slashed{\mathcal{A}}+\slashed{a}\right)\psi_\alpha\right] \,,
}
where $\nabla_\mu$ is the covariant derivative compatible with the background metric $g_{\mu\nu}$, $\psi_\alpha$ are the $N/2$ Dirac fermions\footnote{ Recall that $N$ has to be an even number in order to avoid a parity anomaly in $d=3$. },  and the gauge field has been written as the sum of a background $\mathcal{A}_\mu$, with field strength $\mathcal{F}_{\mu\nu}$, and a small fluctuation $a_\mu$, with field strength $f_{\mu\nu}$, around this background. As usual for a Wilson-Fisher type $\epsilon$ expansion, we have included all terms that are marginal in $4d$, the dimension that we are perturbing away from. 

We are interested in studying this theory on $S^2\times \HH^{2-\epsilon}$ in the background of magnetic flux $4 \pi q$ through $S^2$. We may write the metric as
\es{metric}{
ds^2=R^2_{S}ds^2_{{S^2}}+R^2_{\HH}ds^2_{{\HH^{2-\epsilon}}} \,,
}
where $R_{S}$ and $R_{\HH}$ denote the curvature radii of $S^2$ and $\HH^{2-\epsilon}$, respectively.  We are eventually interested in taking these curvature radii equal, $R_{\HH}= R_{S}$, because that is the case that is related by a conformal transformation to $\R^{4 - \epsilon}$, but for now we will keep them distinct.

In order to have magnetic flux $4\pi q $ through $S^2$, we choose a background gauge field given by:
\es{Amonopoles2}{
\mathcal{A}=q(1-\cos{\theta})d\phi \,, \qquad
 {\cal F} = d {\cal A} = q \sin \theta d\theta \wedge d\phi \,.
}
The expression for ${\cal A}$ is well-defined everywhere away from $\theta=\pi$.  The singularity at $\theta = \pi$ is  not physically observable provided that the Dirac quantization condition $q\in \mathbb{Z}/2$ is obeyed.

Since we will be working with spinors on a curved manifold, we must specify our conventions for the frame and gamma matrices. When applying the $\epsilon$ expansion to spinors, which must be defined in integer dimensions, it is standard practice to use the spinor conventions of the $\epsilon=0$ dimension, i.e. $4d$ in our case. In $4d$, we can write the metric \eqref{metric} explicitly in geodesic polar coordinates as 
\es{metricExplicit}{
ds^2=R^2_{S}(d\theta^2+\sin^2\theta d\phi^2)+R^2_{\HH}(d\eta^2+\sinh^2 \eta d\varphi^2) \,.
}
We then define our frame as 
\es{frame}{
e_1=\partial_\theta \,, \qquad e_2=\frac{1}{\sin\theta}\partial_\phi \,, \qquad e_3=\partial_\eta \,, \qquad e_4=\frac{1}{\sinh \eta}\partial_\varphi \,.
}
We define the gamma matrices on the $4d$ space $S^2 \times \HH^2$ as a tensor product of the standard $2d$ gamma matrices (Pauli matrices $\sigma^1$ and $\sigma^2$) on the separate spaces $S^2$ and $\HH^2$: 
\es{gammas}{
\gamma^1=\sigma^1\otimes1 \,, 
  \qquad
    \gamma^2=\sigma^2\otimes1\,,
    \qquad
    \gamma^3=\sigma^3\otimes\sigma^1\,,
    \qquad
    \gamma^4=\sigma^3\otimes\sigma^2 \,.
}

\subsection{First step:  Setup of the loop expansion}

In the following, we use the $4-\epsilon$ expansion to compute the free energy density 
\es{FreeEnergy}{
\frac{F^{(q)}}{\text{Vol}(\HH^{2-\epsilon})}=-\frac{\log Z^{(q)}}{\text{Vol}(\HH^{2-\epsilon})}
} 
on $S^2 \times \HH^{2-\epsilon} $ in the presence of magnetic flux $4 \pi q$ through $S^2$.\footnote{Because the free energy is dimensionless and we want to obtain a dimensionless number for the density, by $\text{Vol}(M)$, with $M = \HH^{2-\epsilon}$ or $S^2$, we always mean the volume of the unit-radius $M$.}  As explained in the introduction, the first step is to develop a loop expansion \eqref{EpsilonExp0} in terms of $e_0$, and the second step is to find the value of $e_0$ corresponding to the fixed point coupling and plug it back into \eqref{EpsilonExp0}.

The $4-\epsilon$ expansion is a procedure in which we expand every quantity in $\epsilon$ {\it first}, and {\it then} regularize divergences and extract the renormalized quantities. We will see in Section~\ref{largeN} that this order of steps is rather crucial. 

The free energy on $S^2 \times \HH^{2 - \epsilon}$ can be written as
 \es{FreeEnergyEpsilon}{
   \exp\le(-F^{(q)}\ri) &=\exp\le(-\int d^{4-\epsilon}x\ \sqrt{g}\frac{1}{4e_0^2}\mathcal{F}_{\mu\nu} {\cal F}^{\mu\nu} \ri)\\
   &\times
      \int Da\ \exp\le[{N\ov2}\tr\log\left(i\slashed{\nabla}+\slashed{\mathcal{A}}+\slashed{a}\right)_{4-\epsilon} -\int d^{4-\epsilon}x\ \sqrt{g}\frac{1}{4e_0^2}\le(2{\cal F}_{\mu\nu} f^{\mu\nu}+f_{\mu\nu} f^{\mu\nu}\ri) \ri]\,,
 }
where we separated the classical contribution of the background gauge field and integrated out the fermions. Because $\mathcal{A}$ solves the equation of motion of the pure Maxwell theory and hence is a saddle point, the linear in $f$ term above integrates to zero.\footnote{More precisely, we are only considering  gauge field fluctuations that alone have vanishing total flux. Then $a$ is a well defined 1-form, hence $f=da$ is a total derivative and integrates to zero. (Recall that ${\cal F}$ is proportional to the volume form on $S^2$.)}

In $4-\epsilon$ dimensions the theory is weakly coupled and the gauge field fluctuations are small, hence the theory is amenable to a perturbative expansion. The expansion of the functional determinant is
\es{FuncDetExp}{
\tr\log\left(i\slashed{\nabla}+\slashed{\mathcal{A}}+\slashed{a}\right)_{4-\epsilon}&=\tr\log\left(i\slashed{\nabla}+\slashed{\mathcal{A}}\right)_{4-\epsilon} \\
&-\frac{N}{4} \int d^{4-\epsilon}x\, d^{4-\epsilon}x'\ \sqrt{g(x)} \sqrt{g(x')} K^q_{\mu\mu'}(x, x') a^\mu(x)\, a^{\mu'}(x')+\dots
}
where the linear in $a_\mu$ term is absent because $\mathcal{A}$ is a saddle point, and
 \es{KDefMain}{
  K^q_{\mu\mu'}(x,x')  = -\langle J_\mu(x) J_{\mu'}(x') \rangle_{\text{free},4-\epsilon} 
 }
denotes, up to an overall sign, the current-current two-point function in the theory of a free Dirac fermion in the monopole flux background on $S^2 \times \HH^{2-\epsilon}$.    Plugging this expansion back into~\eqref{FreeEnergyEpsilon}, and integrating over the gauge field fluctuations, we find that we can write the free energy as:
 \es{FreeEnergyEpsilonExplicit}{
  F^{(q)} = F_\text{classical}^{(q)} + F_{\text{$1$-loop}}^{(q)} + F_{\text{$2$-loop}}^{(q)} + \cdots \,,
 }
with  
\es{FreeEnergyEpsilon2}{
   F_\text{classical}^{(q)} &\equiv \int d^{4-\epsilon}x\ \sqrt{g}\frac{1}{4e_0^2}\mathcal{F}_{\mu\nu} {\cal F}^{\mu\nu} \,,\\
   F_\text{$1$-loop}^{(q)} &\equiv  -{N\ov2}\tr\log\left(i\slashed{\nabla}+\slashed{\mathcal{A}}\right)_{4-\epsilon}  \,, \\
   F_\text{$2$-loop}^{(q)} &\equiv {N\ov4}\int d^{4-\epsilon}x\, d^{4-\epsilon}x'\  
     \sqrt{g(x)} \sqrt{g(x')} K^q_{\mu\mu'}(x,x')  \Delta_{4-\epsilon}^{\mu\mu'}(x,x')\,,
}
 etc., where $\Delta_{4-\epsilon}^{\mu\mu'}(x,x') = \langle a^\mu(x) a^{\mu'}(x')\rangle_{4-\epsilon}$ is the Maxwell propagator on $S^2\times \HH^{2-\epsilon}$.  The expression \eqref{EpsilonExp0} can then be calculated from \eqref{FreeEnergyEpsilon2}.  

Note that in \eqref{FreeEnergyEpsilon2} we absorbed the $q$-independent functional determinant coming from the term $f^2/4e_0^2$ in  the path integral measure for $a_\mu$, so that it does not show up in~\eqref{FreeEnergyEpsilon2}.  In the following we will explicitly calculate the first two terms  and draw some conclusions about the third term. We emphasize again that these terms should be expanded in $\epsilon$ first, then regularized and renormalized.

\subsection{Classical contribution}\label{background}

The classical contribution of the monopole background to the free energy is at all orders in $\epsilon$ given by
\es{F2freeenergy}{
\frac{F_\text{classical}^{(q)}}{\text{Vol}(\HH^{2-\epsilon})}&=\frac{1}{\text{Vol}(\HH^{2-\epsilon})}\int d^{4-\epsilon}x\sqrt{g}\frac{1}{4e_0^2}\mathcal{F}_{\mu\nu} {\cal F}^{\mu\nu}  =\frac{2\pi R^{2-\epsilon}_{\HH} }{ R_S^2}\, {q^2\ov e_0^2}\,.
}
In the case of $\HH^{2-\epsilon}$, the volume $\text{Vol}(\HH^{2-\epsilon})$ is divergent, but the free energy density in \eqref{F2freeenergy} is finite. The resulting free energy density is dimensionless because $[e_0^2]=\epsilon$ in $4-\epsilon$ dimensions.  For future reference, when $R_{S}= R_\HH= R$, the quantity $f_\text{classical}^{(q)}$ defined in the loop expansion \eqref{EpsilonExp0} is then
 \es{fqClassical}{
  f_\text{classical}^{(q)} = 2\pi q^2 R^{-\epsilon} \,.
 }

The free energy $F^{(q)}/\text{Vol}(\HH^{2-\epsilon})$ interpolates between the scaling dimension $\Delta_q$ of monopole operators in $3d$ and the scaling weight $h_q$ of line operators in $4d$ as defined in~\cite{Kapustin:2005py}.   Indeed, it can easily be checked that the definition in \cite{Kapustin:2005py} agrees with \eqref{F2freeenergy} when $\epsilon = 0$ as follows.  From~\eqref{F2freeenergy}, we have
\es{4ddF}{
\delta F^{(q)}=\left(\frac{q^2}{2e_0^2}\right)\left[\frac{\delta(R^2_{\HH})}{R^2_{S}}-\frac{R^2_{\HH}\, \delta(R^2_{S})}{R^4_{S}}\right]\text{Vol}( S^2 \times \HH^2)\,.
}
Quite generally, the same quantity can be computed from the definition of $T_{\mu\nu}$ as
\es{variation}{
\delta F^{(q)} = -\left\langle {1 \ov 2 } \int d^4x\sqrt{g}\  T_{\mu\nu}\,\delta g^{\mu\nu}\right\rangle \,,
}
where $\delta g^{\mu\nu}$ is the change in the inverse metric when $R_{S}$ and $R_{\HH}$ change by $\delta R_{S}$ and $\delta R_{\HH}$, respectively.  From \cite{Kapustin:2005py} we have that the scaling weight of monopole operators is defined through
\es{hWDef}{
\le\<T_{\mu\nu} \, dx^\mu \otimes dx^\nu\ri\>\equiv -{h_q\ov R^2}\le[ds^2_{S^2}-ds^2_{\HH^{2}}\ri]\,,
} 
where the curvature radii are set to $R^2_{S}=R^2_{\HH}=R^2$, and it equals $h_q=\frac{ q^2}{2e_0^2}$ in pure Maxwell theory.  Plugging this information into \eqref{variation} and setting the curvature radii equal, one immediately reproduces \eqref{4ddF}.\footnote{
In terms of the free energy $F^{(q)}(R_{S},R_{\HH})$ the scaling weight $h_q$ is given by:
\es{h}
{
h_q=\frac{R^2}{\text{Vol}( S^2 \times \HH^2)}\frac{\delta F^{(q)}}{\delta R^2_{S}}\bigg\rvert_{R^2_{S}=R^2_{\HH}=R^2}=-\frac{R^2}{\text{Vol}( S^2 \times \HH^2)}\frac{\delta F^{(q)}}{\delta R^2_{\HH}}\bigg\rvert_{R^2_{S}=R^2_{\HH}=R^2}\,.
}
}

\subsection{One loop contribution at  $O(\epsilon^0)$}\label{determinants}

The next term in~\eqref{FreeEnergyEpsilon2} is the fermionic $1$-loop determinant.  In this section we calculate it to leading order in $\epsilon$, namely $\epsilon^0$, while in Section~\ref{DETEPSILON} we extract some information at next-to-leading order. Hence, corresponding to setting $\epsilon=0$, we want to calculate the following determinant in four dimensions:
\es{freeDets}{
F_\text{$1$-loop}^{(q)}&=-\frac{N}{2}\tr\log\left(i\slashed{\nabla}+\slashed{\mathcal{A}}\right)\,,
}
where the trace is to be computed on the space $ S^2\times \HH^{2}$. We will use zeta function regularization to regulate the divergences in this trace.

We will find that the functional determinant depends on the logarithm of an arbitrary scale $\mu$; the coefficient of $\log\mu$ will be found to be the trace anomaly. Note, however, that $\Delta F^{(q)} \equiv F^{(q)}-F^{(0)}$ is a well-defined quantity in a CFT, so it should be independent of $\mu$.  In Section~\ref{sec:Altogether} we will indeed find that the residual $\log\mu$ dependence, which remains after the subtraction of the vacuum energy $F^{(0)}$, gets absorbed into the expression for the bare coupling at the conformal fixed point order by order in $\epsilon$.  With this preview, let us turn to a brief review of the expected trace anomaly.

\paragraph{Expectation from trace anomaly.}

The trace anomaly for $N$ Weyl fermions is~\cite{Komargodski:2011vj}:\es{stresstraceferm}{
\langle T_\mu^\mu\rangle=&aE_4-cW^2-\beta\le(1\ov e^2\ri)\frac{{\cal F}_{\mu\nu} {\cal F}^{\mu\nu}}{4} \,, \\
a=&\frac{N}{90(8\pi)^2}\frac{11}{2}, \qquad c=\frac{9N}{90(8\pi)^2}, \qquad \beta\le(1\ov e^2\ri)=-\frac{N}{12\pi^2}+\dots\,,
}
where $\beta (1/e^2)$ is the beta function of $1/e^2$.\footnote{The more conventional form of the beta function is
\es{convBeta}{
\beta(e)=-{\epsilon\ov 2}\, e +{N e^3\ov 24\pi^2}+{Ne^5\ov 128 \pi^4}+\dots\,.
}
}
 If we integrate the trace anomaly over $S^2\times \HH^2$, we get the response of the free fermions to the rescaling of $\mu$.\footnote{
 On $S^2\times \HH^2$ the curvature invariants are calculated from the metric~\eqref{metricExplicit} to be
\es{Rs}{
E_4&=R_{\mu\nu\rho\sigma}^2-4R_{\mu\nu}^2+R^2=-\frac{8}{R^2_{S} R^2_{\HH}}\\
W^2&=R_{\mu\nu\rho\sigma}^2-2R_{\mu\nu}^2+\frac{1}{3}R^2=\frac{4}{3}\left({1\ov R^2_{S}}-{1\ov R^{2}_{\HH^2}}\right)^2\\
 {\cal F}^2 &= \frac{2q^2}{R_{S}^4}\,,
}
 in our monopole background.
 }
 Then a standard argument implies that the free energy contains a logarithmic term in $\mu$:
\es{logsferm}{
\frac{F^{(q)}_\text{1-loop}}{\text{Vol}(\HH^{2})}&=-{N\left(-60q^2+3+5r^2+3r^4\right)\ov 360\pi r^2}\log\mu
 + \text{terms independent of $\mu$} \,,
}
where, in order to avoid clutter, we introduced the notation
 \es{rDef}{
  r\equiv \frac{R_{S}}{R_{\HH}} \,.
 }
We will see that our explicit computation of $F^{(q)}_\text{1-loop}$ that we now perform matches the expectation \eqref{logsferm}.

\paragraph{Expansion in modes and functional determinant.}

To calculate the functional determinant, we have to obtain the eigenvalues of the gauge-covariant Dirac operator $i \slashed{\nabla} + \slashed{\cal A}$ and their degeneracies.  Our choice of gamma matrices in \eqref{gammas} implies that this operator can be written as
 \es{Dirac4}{
  i \slashed{\nabla} + \slashed{\cal A} = \frac{1}{R_{S}} \left( i \slashed{\nabla} + \slashed{{\cal A}} \right)_{S^2} \otimes 1 
   + \frac{1}{R_{\HH}} \sigma^3 \otimes i \slashed{\nabla}_{\HH^2} \,,
 }
where $\left( i \slashed{\nabla} + \slashed{{\cal A}} \right)_{S^2}$ is a 2d gauge-covariant Dirac operator (acting on 2-component spinors) on the unit-radius $S^2$ with flux $4 \pi q$ through it, and $i \slashed{\nabla}_{\HH^2}$ is the Dirac operator on the unit-radius $\HH^2$ (also acting on 2-component spinors).  In writing \eqref{Dirac4} we assumed that the 2d gamma matrices on both $S^2$ and $\HH^2$ are $\gamma^1 = \sigma^1$ and $\gamma^2 = \sigma^2$.  The split \eqref{Dirac4} suggests that we should write our 4-component spinors as $\chi \otimes \psi$, where $\chi$ is a 2-component spinor on $S^2$ and $\psi$  is a 2-component spinor on $\HH^2$.

On $S^2$, we can find an orthonormal basis of eigenspinors of $(i \slashed{\nabla} + \slashed{\cal A})$ obeying
 \es{S2Eigen}{
 (i\slashed{\nabla} + \slashed{\cal A} )_{S^{2}} \,  \chi_{q\ell m}^{\left(\pm\right)}=\pm\sqrt{\ell^{2}-q^{2}}\, \chi_{q\ell m}^{\left(\pm\right)} 
    \,,\qquad\chi_{q\ell m}^{\left(\pm\right)}=\sigma_{3}\, \chi_{q\ell m}^{\left(\mp\right)} \,,
  }
with $\ell \geq \abs{q}$ and $-\ell \leq m \leq \ell-1$ such that both $\ell - \abs{q}$ and $\ell - m$ are integers.  While for $\ell>\abs{q}$, the $\chi^{(+)}_{q\ell m}$ and $\chi^{(-)}_{q\ell m}$ are linearly independent, for $\ell = q$ we can identify $\chi^{(+)}_{q, \abs{q}, m}=\chi^{(-)}_{q, \abs{q}, m}\equiv\chi_{q, \abs{q}, m}$---these are the $2 \abs{q}$ zero modes predicted by the Atiyah-Singer index theorem.  See Appendix~\ref{SPINORAPPENDIX} for explicit expressions of these spinor monopole harmonics.  For any given $\ell$, the degeneracy is thus 
 \es{dDef}{
  d_{\ell} = \begin{cases}
   4 \ell \,, & \text{if $\ell > \abs{q}$} \,, \\
   2\abs{q}\,, & \text{if $\ell = \abs{q}$} \,.
  \end{cases}
 }

On $\HH^{2}$, we can find a basis of eigenspinors of $i \slashed{\nabla}_{\HH^2}$ obeying
 \es{H2Eigen}{
    i\slashed{\nabla}_{\HH^{2}}\psi^{(I)}_{\lambda n}=\lambda\psi^{(I)}_{\lambda n} \,,
 }
with $I=1,2$, $\lambda \in \R$, and $n \in \Z_+$.  (There are two linearly independent spinors for each $\lambda$ and $n$.)  Explicit expressions are given in Appendix~\ref{SPINORHARMONICS}.  We denote the  density of states for each set of modes by $\mu^{(I)}_2(\lambda, n)$, which is in fact the same for each mode, so that the combined density of states is
 \es{dosHyp}{
  \mu_2(\lambda,0) = 2\mu^{(I)}_2(\lambda,0) = \frac{\lambda \coth(\pi \lambda) }{2 \pi} \,,
 } 
where we assumed that the $n=0$ modes satisfy $\abs{\psi^{(I)}_{\lambda 0}(0)} = 1$.  We will not need an expression for $\mu_2(\lambda, n)$ for $n > 0$.

On $S^2 \times \HH^2$, then, we can consider the basis of spinors
 \es{S2H2Basis}{
  \xi_{q\ell m\lambda n}^{\left(+,\,I\right)} = \chi_{q\ell m}^{\left(+\right)}\otimes\psi^{(I)}_{\lambda n} \,, \qquad
\xi_{q\ell m\lambda n}^{\left(-,\,I\right)} = \chi_{q\ell m}^{\left(-\right)}\otimes\psi^{(I)}_{\lambda n} \,.
 }
From \eqref{Dirac4}--\eqref{S2H2Basis}, it is easy to see that for $\ell > \abs{q}$, 
 \es{DiracAction}{
    (i\slashed{\nabla} + \slashed{\cal A}) 
      \begin{pmatrix}\xi_{q\ell m\lambda n}^{\left(+,\,I\right)}\\
       \xi_{q\ell m\lambda n}^{\left(-,\,I\right)}
       \end{pmatrix}
   =\begin{pmatrix}{\sqrt{\ell^{2}-q^{2}}/ R_S} &  {\lambda / R_\HH}\\
      {\lambda / R_\HH} & -{\sqrt{\ell^{2}-q^{2}}/ R_S}
       \end{pmatrix}\begin{pmatrix}\xi_{q\ell m\lambda n}^{\left(+,\,I\right)}\\
       \xi_{q\ell m\lambda n}^{\left(-,\,I\right)}
      \end{pmatrix} \,.
 }
When $\ell = \abs{q}$, we have that $\xi_{q,\abs{q}, m\lambda n}^{\left(+,\,I\right)} = \xi_{q,\abs{q}, m\lambda n}^{\left(-,\,I\right)} \equiv \xi^{(I)}_{q,\abs{q}, m\lambda n}$, and
 \es{DiracActionZero}{
    (i\slashed{\nabla} + \slashed{\cal A}) 
     \xi^{(I)}_{q, \abs{q}, m\lambda n}
   =
      \left(\lambda / R_\HH\right) 
       \xi^{(I)}_{q,\abs{q}, m\lambda n} \,.
 }
What \eqref{DiracAction}--\eqref{DiracActionZero} means is that in the basis \eqref{S2H2Basis} the operator $i\slashed{\nabla} + \slashed{\cal A}$ is block diagonal with non-trivial $2\times2$ blocks such as in \eqref{DiracAction}.

Since $S^2 \times \HH^2$ is homogeneous, we have
 \es{trlog}{
  \tr \log (i \slashed{\nabla} + \slashed{\cal A}) = \Vol(S^2) \Vol(\HH^2)\,  \langle x |\log( i \slashed{\nabla} + \slashed{\cal A} )| x \rangle \,,
 }
where $x$ is any point of our choosing.\footnote{We use the normalization
\es{xyNorm}{
\<x\vert y\>={\delta(x-y)\ov \sqrt{g(x)}}\,.
} 
Also recall that $\Vol(M)$ denotes the volume of the manifold $M = S^2$ or $\HH^2$ with unit curvature radius.}  We can take $x$ to correspond to $\theta = \eta = 0$ in the coordinates given in \eqref{metricExplicit}.  With this choice, we can insert a complete set of modes in \eqref{trlog} and find that only the modes with $m = n = 0$ contribute.  The density of these modes is:
\es{lambdaH2}{
  d_\ell \, \mu_2(\lambda,0) =  d_\ell \, {\lambda \coth(\pi\lambda)\ov 2\pi} \,,
 } 
and hence the free energy is given by:
 \es{fermDef}{
f^{(q)}_\text{1-loop}\equiv \frac{F^{(q)}_\text{1-loop}}{\text{Vol}(\HH^{2})}=-\frac{N}{4\pi }\sum_{\ell=q}^\infty d_\ell\int_0^\infty d\lambda \ \lambda \coth(\pi\lambda)\log\left[\frac{\lambda^2}{\mu^2R_\HH^2}+\frac{\ell^2-q^2}{\mu^2R_S^2}\right] \,.
}
Here, $\mu$ is the UV cutoff scale we introduced to make the formulas dimensionally correct, and the argument of the logarithm is just the (absolute value of the) determinant of~\eqref{DiracAction}.   Although $\lambda \in\R$, we only integrate over $\R^+$, which cancels a $1/2$ factor coming from the fact that the determinant of~\eqref{DiracAction} corresponds to the contribution of two modes.

\paragraph{Regularization of the functional determinant.}

 Both the sum and the integral in \eqref{fermDef} are divergent. To regularize them, it is convenient to introduce a slight variation of the usual zeta function regularization procedure \cite{Cognola:2003zt} and write $-\log A$, for some quantity $A$, as
 \es{logIdentity}{
 -\log(A)&=\dds [A^{-s}]\,,
 }
where $\dds$ is a linear functional defined by
 \es{ddsDef}{
 \dds [f] \equiv \lim_{s\to 0} {1\ov n!} {d^n\le(s^{n-1} f(s) \ri)\ov ds^n} \,,
 }
for some positive integer $n$ that we can choose appropriately for our purposes. Standard zeta function regularization corresponds to the choice $n=1$, but higher $n$ choices are needed when the result of the calculation contains singular terms. Acting on a Laurent series, it extracts the coefficient of the linear in $s$ term:
  \es{zetaDemonstration}{
  \dds \le[{a_{-m}\ov s^m}+{a_{-(m-1)}\ov s^{m-1}}+\dots+a_0+a_1 s+\dots\ri]=a_1\,,
 }
 where any $n>m$ in~\eqref{logIdentity} is a good choice. In fact, we will only act with $\dds$ on Laurent series, so~\eqref{zetaDemonstration} can be regarded as the definition of the action of $  \dds$.
 
  We will thus rewrite \eqref{fermDef} as
\es{fermSep}{
f^{(q)}_\text{1-loop}=&\frac{N}{4\pi } \dds \sum_{\ell=q}^\infty d_\ell \left(\underbrace{\int_1^\infty d\lambda \ \lambda \bullet}_{\text{term I}}+\underbrace{\int_0^1 d\lambda \ \lambda \coth(\pi\lambda) \bullet+\int_1^\infty d\lambda \ \lambda \left(\coth(\pi\lambda)-1\right) \bullet}_{\text{term II}}\right)\\
&\times\le(\frac{\lambda^2}{\mu^2R_\HH^2}+\frac{\ell^2-q^2}{\mu^2R_S^2}\ri)^{-s} 
 = \text{I}^{(q)} + \text{II}^{(q)}\,, \\
}
where the integral in the first term (to be referred to as term I) is divergent, while those in the last two terms (term II) are convergent. We have chosen to divide the integration range into $[0,1]\cup[1,\infty)$ so that we can more easily compare with the large $N$ calculation in Section~\ref{largeN}. 

Let us focus on $\text{I}^{(q)}$ first.  After performing the $\lambda$ integral, we have: 
\es{ferm1}{
\text{I}^{(q)}
 &=-\frac{Nq}{4\pi  }(2\log(\mu R_\HH)+1)+\frac{N}{2\pi  }\dds \left[ (\mu R_\HH)^{2s}\sum_{\ell=q+1}^\infty \,\frac{\ell \left(1+\frac{R_\HH^{2}}{R_S^2}(\ell^2-q^2)\right)^{1-s}}{s-1} \right] \,,
 }
 where we have separated out the $\ell=q$ term from the rest.  Next, we add and subtract
 \es{AddSub}{
  r^{2(s-1)}\,\le[{\ell^{3-2s}\ov 1-s}+\le(r^2-q^2\ri)\ell^{1-2s}-{s\le(r^2-q^2\ri)^2\ov 2\ell} \ri]
 }
from each term in the sum in \eqref{ferm1}, where $r \equiv R_{S}/ R_{\HH}$.  The difference between the terms in the sum in \eqref{ferm1} and \eqref{AddSub} results in a convergent sum, while the sum of \eqref{AddSub} itself can be regularized using the Hurwitz zeta function.  This procedure yields a regularized expression for $\text{I}^{(q)}$, which is too lengthy to be reproduced here.  It takes the form
 \es{ferm1Asymp}{
 \text{I}^{(q)} = \frac{N(-1+20q^2+10r^2+30r^4)}{120\pi r^2}\,\log(\mu R_\HH)
  + \text{terms independent of $\mu$}\,.
 }
The full expression when $r=1$ is given in Appendix~\ref{explicit}.

The $\lambda$ integrals in $\text{II}^{(q)}$ are convergent, so we need only regularize the divergent sum over $\ell$, which we do as in $\text{I}^{(q)}$\@.  The answer is again too lengthy to be reproduced here, but it takes the form 
 \es{ferm2Asymp}{
 \text{II}^{(q)} = -\frac{N(35+93r^2)}{360\pi }\,\log(\mu R_\HH)
  + \text{terms independent of $\mu$}\,.
 }
In Appendix~\ref{explicit}, we give the full expression for $\text{II}^{(q)}$ when $r=1$.

Quite nicely, comparing the sum of \eqref{ferm2Asymp} and \eqref{ferm1Asymp} to \eqref{logsferm} we find the expected $\log \mu$ dependence for the functional determinant to this order.   So far, we have kept the radii of $S^2$ and $\HH^2$ different in order to provide a detailed check of our results using the trace anomaly~\eqref{logsferm}.  From now on we will set $R_{S}=R_{\HH}=1$, hence $r=1$.\footnote{The common radius $R_{S}=R_{\HH}=R$ can always be easily reintroduced using dimensional analysis.  By not dropping the $r$-dependence, one could have also extracted some information on integrated $n$-point functions of the stress tensor in the presence of the defect.}

Using the expressions in Appendix~\ref{explicit}, we can write explicitly the quantity $f^{(q)}_\text{1-loop}$ appearing in \eqref{EpsilonExp0} as
 \es{FinalDet}{
 f^{(q)}_\text{1-loop}&=\le(\text{I}^{(q)}+\text{II}^{(q)}\ri)-\le(\text{I}^{(0)}+\text{II}^{(0)}\ri) + O(\epsilon) \\
&\equiv\frac{N\,q^2}{6\pi}\,\log(\mu )+ N \left[ \mathfrak{f}(q) -{ q^2\,\log\le(4 \pi e^{-\gamma}\ri)\ov 12 \pi} \right] + O(\epsilon)\,,
 }
where the second line can be taken as the definition of $ \mathfrak{f}(q)$.  The reason for the slightly awkward definition will become clear in eq.~\eqref{Together2} below.  The quantity $\mathfrak{f}(q)$ is evaluated numerically in Table~\ref{aTable} for a few values of $q$, and its large $q$ behavior is derived in Appendix~\ref{app:largeq}.

\subsection{One loop contribution at $O(\epsilon)$}
\label{DETEPSILON}
 
The computation of the previous section can be easily extended to higher orders in $\epsilon$.  Indeed, the functional determinant can be formally written down in any fractional dimension;  the only difference from the $\epsilon=0$ case is that now the density of states for spinors on $\HH^{2-\epsilon}$ and $n=0$ is \cite{Bytsenko:1994bc}:
\es{NlambdaH2ferm}{
\mu_{2-\epsilon}(\lambda,0)=\frac{1}{2^{3-\epsilon}\pi^{(4-\epsilon)/2} \Gamma\left(\frac{2-\epsilon}{2}\right)}\abs{\frac{\Gamma\left((2-\epsilon)/2+i\lambda\right)\Gamma(i\lambda)}{\Gamma(2i\lambda)}}^2\,,
}
which is a simple generalization of~\eqref{dosHyp}.\footnote{It is easy to check that this is the correct expression, as for $\epsilon=0$ it gives back~\eqref{lambdaH2}, whereas for $\epsilon=1$ it gives $\mu(\lambda)={1\ov \pi}$, which is the appropriate result for three dimensions. In three dimensions we usually write $\int_{-\infty}^\infty {d\omega\ov 2\pi}\ f(\omega^2)$, whereas here we integrate over only the half line, giving $\int_{0}^\infty {d\lambda\ov \pi}\ f(\lambda^2)$.}
Then the functional determinant \eqref{fermDef} for finite $\epsilon$ is 
 \es{FractDimDet}{
\frac{F^{(q)}}{\text{Vol}(\HH^{2-\epsilon})}=-\frac{N}{2}\sum_{\ell=q}^\infty d_\ell\int_0^\infty d\lambda \ \mu_{2-\epsilon}(\lambda,0) \log\left[\frac{\lambda^2}{\mu^2}+\frac{\ell^2-q^2}{\mu^2}\right]
}
where $\mu$ is the UV cutoff.

Expanding the $\HH^{2-\epsilon}$ density of states \eqref{NlambdaH2ferm} to linear order in $\epsilon$, we get
\es{SpectDensLin}{
\mu_{2-\epsilon}(\lambda,0)={\lambda\coth(\pi \lambda)\ov 2\pi} +\le(-{\psi(1+i\lambda)+\psi(1-i\lambda)\ov 2} +{\log\le(4 \pi e^{-\gamma}\ri)\ov 2}\ri)\, {\lambda\coth(\pi \lambda)\ov 2\pi}\epsilon+\dots\,,
}
where $\psi(z)$ is the digamma function.
We see that the second term in the parenthesis gives a contribution to the free energy proportional to the leading term:
\es{2ndTerm}{
\romannumeral2^{(q)} - \romannumeral2^{(0)}  =  N\epsilon\, q^2 \, {\log\le(4 \pi e^{-\gamma}\ri)\ov 12\pi}\, \log \mu
 + \text{terms independent of $\mu$}\,.
}
The contribution from the first term in the parenthesis in \eqref{SpectDensLin} requires a more detailed analysis. Using experience from Section~\ref{determinants}, to determine the $q$-dependent  terms that are proportional to $\log \mu$, we only need to consider the asymptotic behavior (to order $1/\lambda^3$) of the spectral density. In analogy with term I in~\eqref{ferm1} we write:
\es{1stTerm}{
\romannumeral1^{(q)}=& -\frac{N\epsilon}{2\pi}\dds\sum_{\ell=q}^\infty d_\ell\int_1^\infty d\lambda \ \le[\lambda \log\lambda +{1\ov 12 \lambda}+{1\ov 120 \lambda^3}\ri]\left(\frac{\lambda^2+\ell^2-q^2}{\mu^2 }\right)^{-s} \\
=&\frac{N\epsilon \, q}{480\pi }\le(20\log^2\mu+122\log\mu -119\ri)-\frac{N\epsilon}{4\pi }\dds\sum_{\ell=q+1}^\infty \ell\,\mu ^{2s} \,\le({ {}_2 F_1\le(s-1,s-1,s;-(\ell^2-q^2)\ri)\ov (s-1)^2}\ri.\\
&\le.+{{}_2 F_1\le(s,s,s+1;-(\ell^2-q^2)\ri)\ov 6s}+{{}_2 F_1\le(s,s+1,s+2;-(\ell^2-q^2)\ri)\ov 60s}\ri)\,,
}
where the first term is the contribution of $\ell=q$. As in Section~\ref{determinants} we expand for large $\ell$ and use zeta function regularization for the resulting terms.  The remaining finite sum does not give terms proportional to $\log \mu$, so we will not consider it here.  At the end of the day, we have
\es{1stTerm2}{
\romannumeral1^{(q)}-\romannumeral1^{(0)}&=- {N\epsilon\,  q^2\ov 12\pi}\log^2\mu 
 + \text{terms independent of $\mu$}\,,
}
where the linear in $q$ term from the first term in~\eqref{1stTerm} cancels. 

Adding up \eqref{1stTerm} and \eqref{2ndTerm} and copying over the $\epsilon^0$ result in \eqref{FinalDet}, we obtain an improved version of \eqref{FinalDet}:
 \es{FinalDetImproved}{
 f^{(q)}_\text{1-loop}&= \frac{N\,q^2}{6\pi}\,\log(\mu R_\HH)+ N \left[ \mathfrak{f}(q) -{ q^2\,\log\le(4 \pi e^{-\gamma}\ri)\ov 12 \pi} \right] \\
 &{}+ \epsilon \left[- {N q^2\ov 12\pi}\log^2\mu + N q^2 \, {\log\le(4 \pi e^{-\gamma}\ri)\ov 12\pi}\, \log \mu + \text{terms independent of $\mu$} \right] + O(\epsilon^2)\,.
 }
In principle, one can straightforwardly go to higher orders in $\epsilon$ and also compute the terms independent of $\mu$.

 \subsection{Two loop contribution at $O(\epsilon^0)$}\label{sec:current}

\begin{figure}[t]
\begin{center}
\includegraphics[width = .3\textwidth]{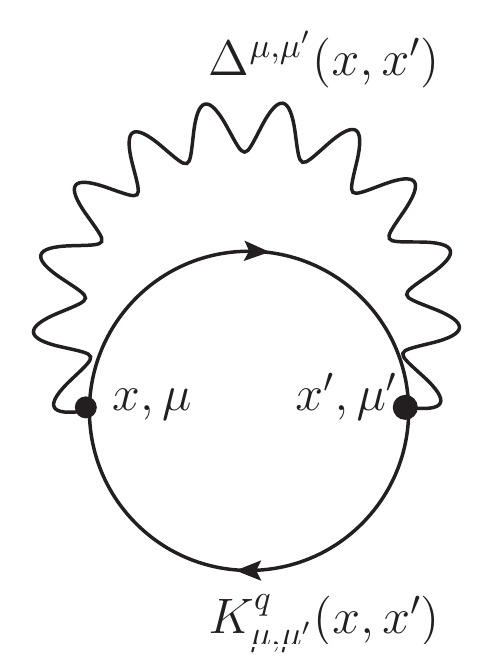}
 \caption{The vacuum Feynman diagram $\mathbb{D}^{(q)}$ that gives the logarithmic contribution~\eqref{Diag3Main} to the free energy.\label{fig:Vacuum}}
 \end{center}
 \end{figure}

Having determined the first two coefficients in \eqref{EpsilonExp} in \eqref{fqClassical} and \eqref{FinalDetImproved} respectively, we can now move on to the two-loop term.  This is the contribution of the current-current correlator to the free energy appearing in the last line of~\eqref{FreeEnergyEpsilon2};  it can be thought of as the vacuum  diagram shown in Figure~\ref{fig:Vacuum}. We will determine its singular piece which is proportional to $\log\mu$. In Appendix~\ref{app:ShortDist} we show that the short distance behavior of this diagram is determined by the flat space limit
 \es{DiagMain}{
  \mathbb{D}^{(q)}- \mathbb{D}^{(0)}=  {N\ov 4} \Vol(S^2 \times \HH^2) \int \frac{d^4 p}{(2 \pi)^4} \  \left[ K^q_{ij}(p) - K^0_{ij} (p)\right] \frac{e_0^2}{p^2} (\delta^{ij} - p^i p^j) \,,
 }
where $i,j$ are frame indices, $p$ is the flat space momentum, and we give formulas for $\left[ K^q_{ij}(p) - K^0_{ij} (p)\right]$ in~\eqref{KMom}.  The $p^i p^j$ term doesn't contribute because $K_{ij}$ is conserved. The first two terms in \eqref{KMom} don't contribute if we use symmetric integration over the momenta.  We're left with the last term:
 \es{Diag3Main}{
  \mathbb{D}^{(q)}- \mathbb{D}^{(0)}=  \Vol(\HH^2)\,   \pi N \int \frac{d^4 p}{(2 \pi)^4}  \ e_0^2  q^2  \frac{2 p_H^2 }{4 \pi^2 \abs{p}^6} \approx  \Vol(\HH^2)\, \frac{N e_0^2 q^2 }{32 \pi^3} \log \mu \,.
 }
We leave the computation of the finite part of $\mathbb{D}^{(q)}$ to future work.

From \eqref{Diag3Main}, we deduce that the coefficient $f_{\text{2-loop}}^{(q)}$ in \eqref{EpsilonExp} is
 \es{f2Loop}{
  f_{\text{2-loop}}^{(q)} = \left[ \frac{N q^2 }{32 \pi^3} \log \mu 
   + \text{terms independent of $\mu$} \right] + O(\epsilon) \,.
 }

\subsection{Determination of fixed point coupling and final result for the free energy}
\label{sec:Altogether}

Putting together \eqref{EpsilonExp}, \eqref{fqClassical}, and \eqref{f2Loop}, eq.~\eqref{EpsilonExp} becomes
 \es{Together}{
\frac{\Delta F^{(q)}}{\text{Vol}(\HH^{2-\epsilon})}
&=\frac{2 \pi q^2 }{e_0^2}
+\Biggl[ \frac{N\,q^2}{6\pi}\,\log(\mu )+ N \mathfrak{f}(q) -{ N q^2\,\log\le(4 \pi e^{-\gamma}\ri)\ov 12 \pi}  \\
 &{}+ \epsilon \left(- {N q^2\ov 12\pi}\log^2(\mu ) + N q^2 \, {\log\le(4 \pi e^{-\gamma}\ri)\ov 12\pi}\, \log (\mu  ) + \text{terms indep.~of $\mu$} \right) + O(\epsilon^2) \Biggr] \\
 &+ e_0^2 \left[ \frac{N q^2 }{32 \pi^3} \log (\mu ) 
   + \text{terms indep.~of $\mu$}+ O(\epsilon) \right]+\dots\,.
 }
By tuning the bare coupling $e_0$ to the weakly coupled conformal fixed point, we should be able to get rid of the $\mu$-dependence of~\eqref{Together}, so that we get an unambiguous answer for the free energy.

We will go through the procedure of determining $e_0$ at the fixed point in some detail to highlight the nontrivial cancellations that occur in the process. The reason why it made sense to further expand each order in the loop expansion in $\epsilon$ is  because the theory is weakly coupled, $e_0^2\sim \epsilon$. Hence we look for $e_0$ in the following form:
\es{RenormalizedCoupling}{
e_0=\alpha\mu^{\epsilon/2} \sqrt{\epsilon}\le(1+c_1\epsilon+\dots\ri)\,.
}
Requiring that the $\log\mu $ dependence cancels at $O(\epsilon^0)$ fixes
\es{alphaValue}{
\alpha=\sqrt{12 \pi^2 \ov N}\,.
}
At $O(\epsilon)$ there is a $\log^2(\mu R)$ term with a coefficient $q^2 \epsilon\le(-\frac{ N}{12 \pi} +\frac{\pi}{\alpha^2} \ri)$. With the value of $\alpha$ given in~\eqref{alphaValue} this term also cancels.  One of the main motivation for the calculations presented in Section~\ref{DETEPSILON} was to see this cancellation. From matching the $\log\mu$ terms at  $O(\epsilon)$ we get that
\es{c1Value}{
c_1=- \frac{9}{8 N } -\frac{\log(4 \pi e^{-\gamma})}{4}\,.
}
We conclude that
\es{RenormalizedCoupling2}{
e_0 &=\le({\mu^2\ov 4 \pi e^{-\gamma} }\ri)^{\epsilon/4}\, \sqrt{12 \pi^2 \epsilon \ov N}\le(1-{9\ov 8 N}\epsilon+\dots\ri)\,,
}
where in the second line we resummed part of $c_1$ into a prefactor familiar from dimensional regularization.\footnote{In dimensional regularization logarithmic divergent integrals give $1/\epsilon$ poles that are accompanied by the prefactor in~\eqref{RenormalizedCoupling2}:
\es{DimRegUsual}{
\text{log.~div.~int.}={1\ov \epsilon}+\frac12 \log\le(\mu^2\ov 4 \pi e^{-\gamma}\ri)+\text{finite}\,.
} } 

Using the fixed point value of $e_0$,~\eqref{Together} becomes:
\es{Together2}{
\frac{\Delta F^{(q)}}{\text{Vol}(\HH^{2-\epsilon})}
=&\frac{N q^2}{6\pi \epsilon  }+N {\mathfrak f}(q) +{3q^2\ov 8 \pi}+O(\epsilon)\,.
 }
This concludes the calculation in the $\epsilon$ expansion. Before turning our attention to two consistency checks in the large $q$ limit and in the large $N$ limit of \eqref{Together2}, respectively, we digress to derive the fixed point coupling constant~\eqref{RenormalizedCoupling2} from another point of view.

\subsubsection{More conventional derivation of the fixed point coupling}

The expression \eqref{RenormalizedCoupling2} for the fixed point coupling can also be understood from a more conventional point of view. Note that by computing the free energy in a flux background we have essentially performed the computation of the effective potential of the gauge field. Gauge invariance dictates that the effective potential can only depend on $q$. We have computed the one-loop contribution exactly and have determined the divergent piece of the two-loop term. It is then standard procedure to read off the two-loop beta function from such a calculation. Because our zeta function scheme is such that it doesn't produce any divergences at any order of perturbation theory (unlike dimensional regularization), we don't get $1/s$ poles at the end of the calculation, and in our scheme there is a very simple relation between the bare and renormalized coupling
\es{ERDef}{
e_R(\mu)&\equiv \le({\mu^2\ov 4 \pi e^{-\gamma} }\ri)^{-\epsilon/4}\, e_0\,.
}
We chose to absorb the $4\pi e^{-\gamma}$ term into this definition; this prescription is reminiscent of what one does in the $\overline{\text{MS}}$ scheme. This choice is not necessary, as we discuss below.  The $\beta$-function of the renormalized coupling then can be determined from requiring $F^{(q)}$ to be independent of $\mu$. We get
\es{betas}{
\beta(e_R)&=-\frac{e_R}{2}\epsilon+\frac{Ne_R^3}{24\pi^2}+\frac{Ne_R^5}{128\pi^4}+O(e_R^7)\,,
}
which results at a weakly coupled fixed point at
\es{FixedER}{
e_R=\sqrt{12 \pi^2 \epsilon \ov N}\le(1-{9\ov 8 N}\epsilon+O(\epsilon^2)\ri)\,.
}
Putting~\eqref{ERDef} and~\eqref{FixedER} together, we arrive at the result~\eqref{RenormalizedCoupling2}. Had we not chosen to include the $4\pi e^{-\gamma}$ factor in~\eqref{ERDef}, we would have gotten a different result for~\eqref{betas} at $O(\epsilon)$.\footnote{The well-known statement of scheme-independence of the beta function up to two loops only refers to the $O(\epsilon^0)$ piece of the beta function.}  Upon including the  $4\pi e^{-\gamma}$ factor in~\eqref{ERDef}, however, our beta function~\eqref{betas} agrees with that computed in the MS scheme in dimensional regularization~\cite{Gorishnii:1991hw}.

\subsection{The large $q$ limit}

The first check of our final expression \eqref{Together2} comes from its large $q$ behavior, which was also mentioned in the introduction. Following the techniques introduced in~\cite{Dyer:2015zha}, we calculate the large $q$ behavior of the functional determinant in Appendix~\ref{app:largeq} to order $q^{-1}$. The leading order result is: 
\es{largqFinalFermMain}{
 {\mathfrak f}(q)&=-\frac{q^2}{12\pi}\log\left(\frac{q}{2 \pi e^{-\gamma} A^{12}}\right)+O(q)\,,
}
where $A$ is the Glaisher constant.
From~\eqref{Together2}, we then obtain the large $q$ behavior of the free energy, which can be written as
\es{largeqFull}{
\frac{\Delta F^{(q)}}{\text{Vol}(\HH^{2-\epsilon})}
= {\cal N}(\epsilon) \, q^2 \le[1-{\epsilon\ov 2} \log\left(q\right)+O\le({1\ov q},\epsilon^2\ri)\ri] \,,
}
where we introduced
\es{calN}{
 {\cal N}(\epsilon)\equiv {N \ov 6\pi\epsilon}\le[1+ \le(\frac12 \log\left(2 \pi e^{-\gamma}\, A^{12}\right)+{9\ov 4N}\ri)\epsilon+O(\epsilon^2)\ri]\,.
}
Quite remarkably, this expression can be resummed into
 \es{largeqResum}{
  \frac{\Delta F^{(q)}}{\text{Vol}(\HH^{2-\epsilon})} 
   = {\cal N}(\epsilon) \, q^{(4-\epsilon)/2}+O\le(q,\epsilon\ri)\,,
 }
which matches the expectation~\eqref{Largeq}.  

\begin{figure}[!h]
\begin{center}
\includegraphics[width = 0.7\textwidth]{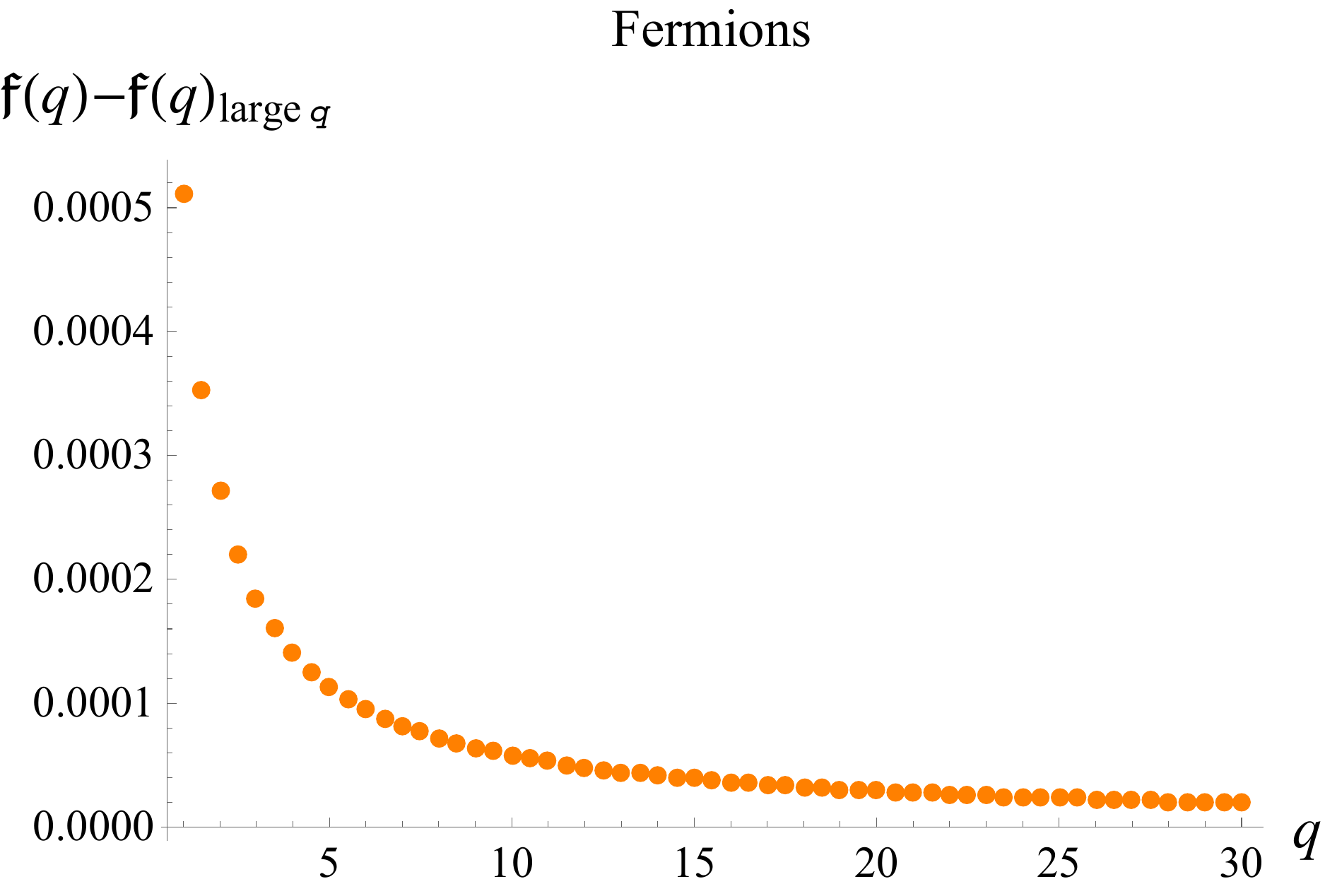}
 \caption{This plot shows the difference between the large $q$ approximation~\eqref{largqFinalFerm} and the arbitrary $q$ numerics (with the exact $q=0$ part subtracted). \label{fig:Compare}}
 \end{center}
 \end{figure}

In Figure~\ref{fig:Compare} we compare the analytic large $q$ expansion to order $O(q^{-1})$ determinant \eqref{largqFinalFerm} to the exact numerical expression. In this comparison we subtract the $q=0$ part of the exact numerical free energy from the analytic large $q$ expansion so that the comparison is accurate. Note that the plot goes like $q^{-1}$ and asymptotes to zero, as we would expect for a large $q$ approximation to order $O(q^{-1})$. The large $q$ expansion seems to be extremely accurate even for small $q$.  We see that even for the smallest possible value of $q =1/2$, the error is only $O(10^{-4})$.

 \section{The large $N$ expansion}\label{largeN}
 
As already mentioned, another perturbative approach to the theories of interest is the large $N$ expansion.   In non-supersymmetric theories, this method is pretty much the only way monopole operators in $3d$ gauge theories have been studied thus far.  
 
 The large $N$ expansion can be performed in any space-time dimension $d = 4-\epsilon$.  As mentioned in the introduction, it has the structure given in \eqref{NExp}, which we reproduce here for the reader's convenience:
\es{NExp2}{
{\Delta F^{(q)}\ov \Vol\le(\HH^{2-\epsilon}\ri) }=N_{} \, g_{0}^{(q)}(\epsilon)+ g_{1}^{(q)}(\epsilon)+ \frac{g_{2}^{(q)}(\epsilon) }{N} + \dots\,.
}
In this section, we will determine $g_{0}^{(q)}(\epsilon)$ and compare the $\epsilon \to 0$ limit of the result to the $4-\epsilon$ expansion result of the previous section.

An interesting subtlety is that, when $\epsilon > 0$, the Maxwell term is irrelevant, and hence it shouldn't be included in the Lagrangian if we wish to describe the IR CFT\@.  Consequently, in contrast to the $4-\epsilon$ expansion method, in the large $N$ approach we do not have a tunable gauge coupling $e_0$; instead, the theory is automatically conformally invariant.  It is thus quite non-trivial to see how the results from the $4-\epsilon$ expansion agree with those from the large $N$ expansion in their overlapping regime of validity.

Even though there is no Maxwell term in the action, the gauge field nevertheless acquires a non-local kinetic term from quantum effects due to the massless matter fields;  since at large $N$ there are many matter fields, the gauge field fluctuations are suppressed and the theory is perturbative. The leading contribution to the free energy comes from the matter field fluctuations around the monopole background. At subleading order, the Gaussian gauge field fluctuations contribute to give:
\es{LargeNSketch}{
{F^{(q)}\ov \Vol\le(\HH^{2-\epsilon}\ri) }=-{N\ov 2} \tr\log \le(i\slashed{\nabla}_{4-\epsilon}\ri)+\frac12\,``\tr\log K"+\dots\,,
}
where $\slashed{\nabla}_{4-\epsilon}$ is the background gauge covariant derivative and $K$ is (up to contact terms) the current-current correlation function of the CFT~\eqref{KDefMain}, which becomes the kernel of the gauge field fluctuations. By comparing~\eqref{NExp2} with~\eqref{LargeNSketch}, we see that $g_{0}^{(q)}(\epsilon)$ is determined from the functional determinant, and $g_{1}^{(q)}(\epsilon)$ is calculable from $K$. We put $\tr\log K$ in quotation marks in~\eqref{LargeNSketch} to indicate that there are subtleties related to gauge invariance that will not be important here. Similar calculations were performed in three-dimensional theories with fermionic matter in~\cite{Pufu:2013vpa,Dyer:2013fja} and in bosonic theories~\cite{Pufu:2013eda,Dyer:2015zha}. 
 
In Sections \ref{background} and \ref{determinants} we calculated the $O(1/\epsilon)$ and $O(\epsilon^0)$ corrections to the free energy by first expanding the action \eqref{firstAction} for small $\epsilon$, and then regularizing the resulting terms.   In terms of the regularization parameter $s$, that method involved taking $\epsilon \to 0$ first and then $s \to 0$.

By contrast, in the large $N$ method we first take $s \to 0$ and then, if we wish to describe the theory close to four dimensions, we can take $\epsilon \to 0$.  Indeed, expanding the regularized functional determinant at small $\epsilon$ and comparing to the relevant terms from the $\epsilon$ expansion method will allow us to check the results of the $\epsilon$ expansion method.   In some sense, the large $N$ results are interesting in that they provide an orthogonal viewpoint on our computation.

\subsection{Functional determinant at finite $\epsilon$}

 The generalization of the free energy \eqref{fermDef} to finite $\epsilon$ is given in~\eqref{FractDimDet} in the case where the curvature radii are $R_{S}= R_{\HH} = 1$.  For general $R_{S}$ and $R_{\HH}$, it takes the form
 \es{largeNfermDef}{
\frac{F^{(q)}}{\text{Vol}(\HH^{2-\epsilon})}=-\frac{N}{2}\sum_{\ell=q}^\infty d_\ell\int_0^\infty d\lambda \ \mu_{2-\epsilon}(\lambda,0) \log\left[\frac{\lambda^2}{\mu^2R_{\HH}^2}+\frac{\ell^2-q^2}{\mu^2R_S^2}\right]\,,
}
where the density of states $\mu_{2-\epsilon}(\lambda, 0)$ was given in \eqref{NlambdaH2ferm} and the degeneracy of the modes on $S^2$ was given in \eqref{dDef}.

\begin{figure}[!h]
\begin{center}
\includegraphics[width = 0.7\textwidth]{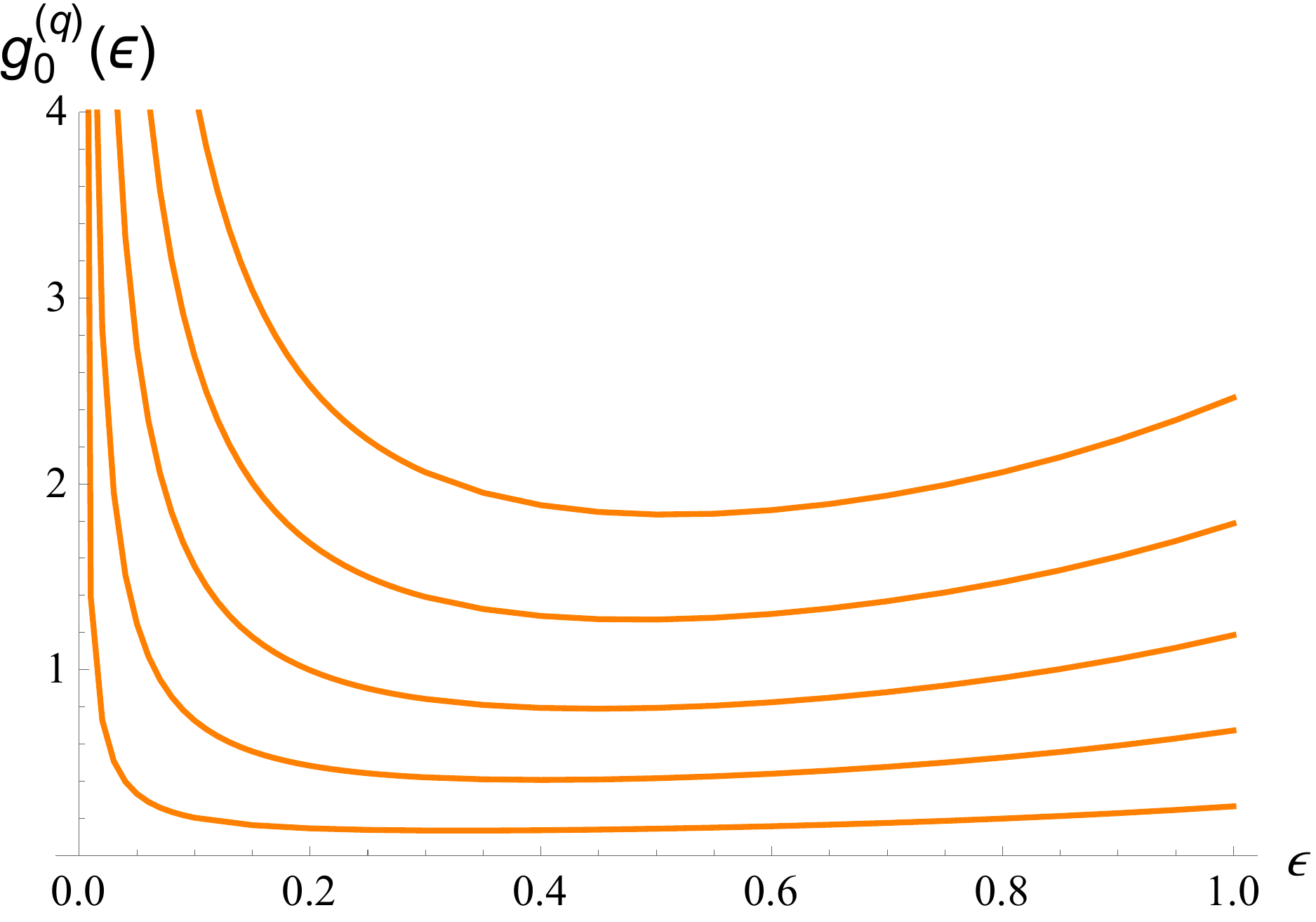}
 \caption{Large $N$ finite $\epsilon$ functional determinant for $q=1/2,1,3/2,2,5/2$.   \label{fig:determinant}}
 \end{center}
 \end{figure}

The method for computing \eqref{largeNfermDef} is as in Section~\ref{determinants}.  In particular, we first divide the integral into the large $\lambda$ divergent part and the remaining convergent part, thus writing 
 \es{FSum}{
  \frac{F^{(q)}}{\text{Vol}(\HH^{2-\epsilon})}={\cal I}^{(q)} + {\cal II}^{(q)} \,,
 }
with
  \es{termILargeN}{
  {\cal I}^{(q)}=&- \frac{N}{2 }\dds\Biggl[\sum_{\ell=q}^\infty d_\ell \int_1^\infty d\lambda \ \mu^{\text{asymp}}_{2-\epsilon}(\lambda) \left(\frac{\lambda^2}{\mu^2 R_\HH^2}+\frac{\ell^2-q^2}{\mu^2R_S^2} \right)^{-s}\Biggr] \,, \\
  {\cal II}^{(q)}=&- \frac{N}{2 }\dds \Biggl[ \sum_{\ell=q}^\infty d_\ell \le(\int_0^1 d\lambda \ \mu_{2-\epsilon}(\lambda)\bullet+\int_1^\infty d\lambda \ \le(\mu_{2-\epsilon}(\lambda)-\mu^{\text{asymp}}_{2-\epsilon}(\lambda)\ri)\bullet\ri)\\
   &\times\left(\frac{\lambda^2}{\mu^2 R_\HH^2}+\frac{\ell^2-q^2}{\mu^2R_S^2} \right)^{-s} \Biggr] \,.
}
Here, 
 \es{muAsympDef}{
  \mu^{\text{asymp}}_{2-\epsilon} \equiv&\frac{2^{\epsilon -1} \pi ^{\frac{\epsilon }{2}-1} }{ \Gamma \left(1-\frac{\epsilon }{2}\right)}\left[  \lambda^{1 - \epsilon}
    -
   \lambda^{-1-\epsilon} \frac{\Gamma(\epsilon + 1)}{24\, \Gamma(\epsilon - 2)} 
   + \lambda^{-3 - \epsilon} \frac{ (5 \epsilon - 12) \Gamma(\epsilon + 3)}{5760\, \Gamma(\epsilon - 2)}
   \right] 
 }  
is just an approximation of $\mu_{2-\epsilon}(\lambda)$ at large $\lambda$, and $s$ is a regularization parameter.  Then, we use zeta function regularization to regularize the divergent sums and integrals.  (Note that now that $\epsilon$ is nonzero, term ${\cal I}^{(q)}$ would have been infrared divergent had the lower limit of integration been $0$;  to avoid this problem, we divided the integration range into $[0,1]\cup[1,\infty)$.\footnote{This division was unnecessary in~\eqref{fermSep} in Section~\ref{determinants}.  We nevertheless performed it there too only to facilitate the comparison with the results in this section.})  The final expressions are rather complicated, and we will not reproduce them here.  We plot the final result in Figure~\ref{fig:determinant} as a function of $\epsilon$.

Let us now explain how the $\epsilon \to 0$ limit of this large $N$ result is compatible with the small $\epsilon$ expansion previously obtained.  This limit should be taken carefully, because the $\epsilon \to 0$ limit does not commute with the $s \to 0$ limit that is implicit in the definition of the linear functional $\dds$.  Actually, the two limits do not commute only for ${\cal I}^{(q)}$;  they commute for ${\cal II}^{(q)}$, and thus
  \es{ferm2AsympLargeN}{
  {\cal II}^{(q)} = \left[ -\frac{N(35+93r^2)}{360\pi }\,\log(\mu R_\HH)
  + \text{terms independent of $\mu$} \right] + O(\epsilon)\,,
 }
with the $O(\epsilon^0)$ term matching \eqref{ferm2Asymp}.  Here, $r \equiv R_{S}/ R_\HH$, just as before.  The expression for ${\cal I}^{(q)}$ is rather complicated and we do not reproduce it in its entirety here.  It is of the form: 
 \es{ferm1LargeN}{
{\cal I}^{(q)}=&-\frac{N(-60q^2+3+5r^2+3r^4)}{360\pi r^2}\,{1\ov \epsilon}+ \Biggl[ \frac{N(35+93r^2)}{360\pi }\log(\mu R_\HH)\\
&+\text{terms independent of $\mu$} \Biggr] + O(\epsilon)\,.
 }
It differs from $\text{I}^{(q)}$ in \eqref{ferm1Asymp} in that it starts at order $1/\epsilon$ and in that the order $\epsilon^0$ terms, in particular the $\log \mu$ dependence, are also slightly different.

 There are two important aspects of this result. First, in fractional dimensions we do not expect to see any trace anomaly, and correspondingly the free energy should be independent of $\mu$. Indeed, the $\log\mu$ terms cancel between~\eqref{ferm1LargeN} and~\eqref{ferm2Asymp}, which is different from what happened when we worked in the $\epsilon$ expansion in the previous section.  Second, we now encounter terms proportional to $1/\epsilon$, the coefficient of which is exactly the trace anomaly~\eqref{logsferm}. 
 
Setting $R_{S}= R_\HH = 1$ from now on, we find\footnote{More generally, we expect that the difference between ${\cal I}^{(q)} - {\cal I}^{(0)}$ and the corresponding quantity computed by expanding in $\epsilon$ first and then regularizing resums to $\frac{N  q^2 (4 \pi e^{-\gamma})^{\epsilon /2}} { 6 \pi \mu^\epsilon \epsilon } $.  In \eqref{termIdiffGeneral} we see only the first two terms in the small $\epsilon$ expansion of this quantity.}
 \es{termIdiffGeneral}{
\le[({\cal I}^{(q)}  - \text{I}^{(q)})-({\cal I}^{(0)}-\text{I}^{(0)})\ri] &=
 \frac{N q^2}{6\pi}\,{1\ov \epsilon} + {Nq^2\ov 12 \pi}\le[ \log (4\pi e^{-\gamma})-2\log \mu  \ri] + O(\epsilon) \,, \\
 \le[({\cal II}^{(q)}  - \text{II}^{(q)})-({\cal II}^{(0)}-\text{II}^{(0)})\ri] &=
  O(\epsilon) \,.
}
In \eqref{FinalDet}, we found
 \es{FinalDetLargeN}{
   \le(\text{I}^{(q)}+\text{II}^{(q)}\ri)-\le(\text{I}^{(0)}+\text{II}^{(0)}\ri)  = \frac{N\,q^2}{6\pi}\,\log \mu+ N \left[ \mathfrak{f}(q) -{ q^2\,\log\le(4 \pi e^{-\gamma}\ri)\ov 12 \pi} \right] + O(\epsilon)\,.
 }
Adding up \eqref{termIdiffGeneral}--\eqref{FinalDetLargeN}, we obtain
 \es{FinalcalSum}{
   \frac{\Delta F^{(q)}}{\text{Vol}(\HH^{2-\epsilon})} \approx \le({\cal I}^{(q)}+{\cal II}^{(q)}\ri)-\le({\cal I}^{(0)}+{\cal II}^{(0)}\ri)
      = N \left[ \frac{ q^2}{6\pi}\,{1\ov \epsilon} 
     +   \mathfrak{f}(q) + O(\epsilon) \right] \,,
 }
which at large $N$ agrees precisely with the expression \eqref{Together2} we obtained in the $\epsilon$ expansion!

It is quite remarkable how the fermion determinant computed in the order of limits $s \to 0$ then $\epsilon \to 0$, reproduces, for instance, the free energy contribution from the Maxwell action evaluated on the classical background when the coupling constant $e_0$ is tuned to its critical value.\footnote{It is worth noting that terms such as the $\log (4\pi e^{-\gamma})$ appearing \eqref{termIdiffGeneral} were also encountered in~\cite{Hawking:1976ja} when comparing the results of zeta function and dimensional regularizations on product manifolds.}  For $\epsilon > 0$, there was no coupling constant that needed to be tuned to criticality, and yet the result is continuously connected to that in the $4-\epsilon$ expansion where such a coupling was explicit in the action.  We therefore regard the computation in this section as a highly non-trivial check of the approach presented in the previous section.

In the Introduction we argued that from the consistency between the large $N$ and $\epsilon$ expansions, the coefficients $g_0^{(q)}$ and $g_1^{(q)}$ in~\eqref{NExp2} should have the $\epsilon$ expansion given in~\eqref{Consistency}.
 We have thus confirmed the expectation~\eqref{Consistency} for $g_0^{(q)}$ directly from a large $N$ computation.  The expectation for $g_1^{(q)}$ should follow from a similar computation of the gauge field fluctuations, where again the order in which we take $\epsilon \to 0$ and remove the regulator should make a difference.

\section{Remarks on the flavor symmetry transformation of defect operators}\label{sec:global}

As discussed in detail in~\cite{DiPietro:2015taa}, $d=4$ QED with $N/2$ flavors has an $SU(N/2)_L\times SU(N/2)_R$ flavor symmetry, where $L$ rotates only the left-handed components of the fermions and $R$ only the right-handed ones.  In the $4-\epsilon$ expansion only the diagonal combination survives because of the impossibility of continuing $\gamma_5$ to fractional dimensions while preserving all the other symmetries of the theory.   In $d=3$,  the flavor symmetry gets enhanced to $SU(N)$.\footnote{It is commonly believed that for sufficiently small $N$, $SU(N)$ is spontaneously broken to a subgroup by dynamical fermion masses.  Our computation of the conformal dimensions of monopole operators could potentially be used to make the case for such breaking, though our results do not apply if the symmetry is broken.}  It is thus natural to ask if the defect operators studied in this paper transform under the flavor symmetry. 

In $d=3$ it is known how the monopole operators transform under the flavor symmetry \cite{Borokhov:2002ib, Dyer:2013fja}.  The Dirac operator on $S^2$ in the presence of monopole flux has $2 \abs{q}$ zero modes.  From the point of view of the field theory on $S^2 \times \R$, these zero modes yield $2 \abs{q}$ zero-energy modes for each flavor.  The corresponding creation operators can act on the vacuum and generate a $2^{2 N \abs{q}}$ Fock space.  However, since the fermions carry gauge charge, not all these states are gauge invariant.  The Gauss law constraint and CP invariance restrict the physical vacua to be Lorentz scalars and to transform under $SU(N)$ as the irreducible representation with the Young diagram \cite{Borokhov:2002ib, Dyer:2013fja}
 \es{SUNfRep}{
  {\tiny N/2} \Bigl\{   \underbrace{ {\tiny \ydiagram{3, 3}} }_{2 \abs{q}} \,.
 } 

It would be interesting to generalize the above discussion to $d = 4 - \epsilon$ dimensions.  On $S^2 \times \HH^2$, as can be seen from~\eqref{DiracActionZero} by setting $\lambda=0$, there are many zero modes of the Dirac operator in the presence of monopole flux through $S^2$ of the form $ \xi^{(I)}_{q,\abs{q}, m,0n} =\chi_{q,\abs{q}, m} \otimes \psi_{0n}^{(I)} $, with $\psi_{0n}^{(I)}$ defined in \eqref{psiHypDefs} and $\chi_{q,\abs{q}, m}$ defined in \eqref{chipmDef}.  (When $\ell = \abs{q}$, $\chi^{(+)}_{q, \abs{q}, m}=\chi^{(-)}_{q, \abs{q}, m}\equiv\chi_{q, \abs{q}, m}$.)  The modes with $I=1$ transform only under $SU(N/2)_L$ while those with $I=2$ transform only under $SU(N/2)_R$.  We leave the analysis of how these zero modes induce the transformation properties of the defect operators under the flavor symmetries to future work.  In particular, it would be interesting to see how defect operators transforming in different irreducible components of the flavor symmetry might acquire different scaling weights.  We expect that such a difference, if present, would appear in the two-loop computation that we have not carried out here.

\section*{Acknowledgments}

We thank Davide Gaiotto, Simone Giombi, Igor Klebanov, and Grisha Tarnopolsky for useful discussions.  The work of SMC and SSP was supported in part by the US NSF under grant No.~PHY-1418069.  MM was supported by the Princeton Center for Theoretical Science. The work of IY was supported in part by the US NSF under grant No.~PHY-1314198 and in part by World Premier International Research Center Initiative (WPI), MEXT, Japan.

\appendix

 \section{Explicit expressions for the modes} \label{app:explicit}

 \subsection{Spinor harmonics on $\HH^2$}
\label{SPINORHARMONICS}

The eigenfunctions of the Dirac operator on hyperbolic space were
computed in \cite{Camporesi:1995fb}. We have
 \es{psiHypDefs}{
    \psi_{\lambda n}^{\left(1\right)}
       =
        \begin{pmatrix}\left(\cosh\frac{\eta}{2}\right)^{n+1}\left(\sinh\frac{\eta}{2}\right)^{n}\,_{2}\text{F}_{1}\left(1+n+i\lambda,1+n-i\lambda,n+1,-\sinh^{2}\frac{\eta}{2}\right)\\
           -\frac{i\lambda}{n+1}\left(\cosh\frac{\eta}{2}\right)^{n}\left(\sinh\frac{\eta}{2}\right)^{n+1}\,_{2}\text{F}_{1}\left(1+n+i\lambda,1+n-i\lambda,n+2,-\sinh^{2}\frac{\eta}{2}\right)
        \end{pmatrix}   
          e^{i\left(n+\frac{1}{2}\right)\varphi} \,, \\
    \psi_{\lambda n}^{\left(2\right)}
       =
        \begin{pmatrix} - \frac{i\lambda}{n+1}\left(\cosh\frac{\eta}{2}\right)^{n}\left(\sinh\frac{\eta}{2}\right)^{n+1}\,_{2}\text{F}_{1}\left(1+n+i\lambda,1+n-i\lambda,n+2,-\sinh^{2}\frac{\eta}{2}\right)\\
           \left(\cosh\frac{\eta}{2}\right)^{n+1}\left(\sinh\frac{\eta}{2}\right)^{n}\,_{2}\text{F}_{1}\left(1+n+i\lambda,1+n-i\lambda,n+1,-\sinh^{2}\frac{\eta}{2}\right)
        \end{pmatrix}
         e^{-i\left(n+\frac{1}{2}\right)\varphi} \,,
  }
such that 
 \es{DiracH2}{
  i\slashed{\nabla}_{\HH^{2}}\psi_{\lambda n}^{\left(I\right)}= \lambda\psi_{\lambda n}^{\left(I\right)}\,, \qquad I=1,2 \,.
 }
Let us denote the density of these modes by $\mu_2^{(I)} (\lambda, n)$.  It is defined by
 \es{muGenDef}{
  \left(\psi_{\lambda n}^{(I)}, \psi_{\lambda' n'}^{(I')} \right) \equiv \int d \eta\, d\phi\, \sinh \eta\,  \le(\psi_{\lambda n}^{(I)}\ri)^\dagger \psi_{\lambda' n'}^{(I')}= \frac{1}{\mu_2^{(I)} (\lambda, n) } \delta(\lambda - \lambda') \delta^{II'} \delta_{n n'} \,.
   }
For the spherical modes (namely those with $n=0$), we have 
 \es{mu2Def}{
   \mu^{(I)}_2\left(\lambda, 0\right)=\frac 1{4 \pi} \lambda\coth\pi\lambda \,.
 }

\subsection{Spinor monopole harmonics}
\label{SPINORAPPENDIX}

The spinor monopole spherical harmonics can be described using the usual scalar monopole spherical harmonics $_{q}Y_{\ell m}\left(\theta,\phi\right)$, as explained in \cite{Borokhov:2002ib, Pufu:2013vpa}.  A minor difficulty is that in these references, the spinor harmonics were written in a basis for the spin bundle which is the result of a conformal transformation from $\mathbb{R}^{3}$ to $\mathbb{R}\times S^{2}$.  If $\tau$ parametrizes $\R$, the frame used in \cite{Borokhov:2002ib, Pufu:2013vpa} can be written as
 \es{eCart}{
   \tilde e^{1}&=\cos\theta\cos\phi d\theta-\sin\theta\sin\phi d\phi+\sin\theta\cos\phi d\tau \,, \\
   \tilde e^{2}&=\cos\theta\sin\phi d\theta+\sin\theta\cos\phi d\phi+\sin\theta\sin\phi d\tau \,, \\
   \tilde e^{3}&=-\sin\theta d\theta+\cos\theta d\tau \,.
 }
In order to borrow their results, we should perform a frame change to
 \es{eS2R}{
  e^{1}&=d \theta \,, \\
  e^{2}&=\sin \theta d\phi \,, \\
  e^{3}&=d\tau \,,
 }
and then simply erase the $\R$ direction along with $e^{3}$.  To match our conventions, we should also take $\gamma_i = \sigma_i$, as was also done in \cite{Borokhov:2002ib, Pufu:2013vpa}.

The two frames \eqref{eCart} and \eqref{eS2R} differ by a frame rotation.  If we write 
 \es{Re}{
  R^a{}_b \tilde e^{b}{}_\mu = e^a{}_\mu \,,
 }
we can identify ${R^{a}}_{b}\left(\theta,\phi\right)={e^{a}}_{\mu} \tilde e^{\mu}{}_{b}$.  Acting on spinors, the matrix $R$ takes the form
 \es{GotR}{
  R = \begin{pmatrix}
   e^{i \phi / 2} \cos \frac{\theta}{2} & e^{-i \phi / 2} \sin \frac{\theta}{2} \\
   -e^{i \phi/2} \sin \frac{\theta}{2} & e^{-i \phi /2 } \cos \frac{\theta}{2} 
  \end{pmatrix} \,.
 }
and thus a spinor $\chi$ written in the frame \eqref{eS2R} is related to a spinor $\tilde \chi$ in the frame \eqref{eCart} by
 \es{SpinorTransf}{
  \chi = R \tilde \chi \,.
 }

In the conventions in the main text, the basis described in \cite{Pufu:2013vpa} is then
 \es{TSDef}{
   T_{q,\ell m}\left(\theta,\phi\right) &= 
    R  \left(\begin{array}{c}
     \sqrt{\frac{\ell+m+1}{2\ell+1}}\,_{q}Y_{\ell m}\left(\theta,\phi\right)\\
     \sqrt{\frac{\ell-m}{2\ell+1}}\,_{q}Y_{\ell\, m+1}\left(\theta,\phi\right)
     \end{array}\right) \,, \\
  S_{q,\ell m}\left(\theta,\phi\right) &=
    R  \left(\begin{array}{c}
     -\sqrt{\frac{\ell-m}{2\ell+1}}\,_{q}Y_{\ell m}\left(\theta,\phi\right)\\
     \sqrt{\frac{\ell+m+1}{2\ell+1}}\,_{q}Y_{\ell\, m+1}\left(\theta,\phi\right)
     \end{array}\right) \,.
 }
On the basis spinors, the gauge-covariant Dirac operator in a monopole background acts as
 \es{DiracTS}{
i\slashed D\left(\begin{array}{c}
T_{q,\ell-1\,m}\\
S_{q,\ell m}
\end{array}\right)=\left(\begin{array}{cc}
0 & i\sqrt{\ell^{2}-q^{2}}\\
-i\sqrt{\ell^{2}-q^{2}} & 0
\end{array}\right)\left(\begin{array}{c}
T_{q,\ell-1\,m}\\
S_{q,\ell m}
\end{array}\right),
 }
such that the eigenvector/eigenvalue combinations are 
 \es{chipmDef}{
    \chi_{q\ell m}^{\left(+\right)} &\equiv \frac{-1}{\sqrt{2}}e^{-i\alpha}
      R \left(\begin{array}{c}
       -i\sqrt{\frac{\ell+m}{2\ell-1}}\,_{q}Y_{\ell-1\,m}\left(\theta,\phi\right)-\sqrt{\frac{\ell-m}{2\ell+1}}\,_{q}Y_{\ell m}\left(\theta,\phi\right)\\
       -i\sqrt{\frac{\ell-m-1}{2\ell-1}}\,_{q}Y_{\ell-1\,m+1}\left(\theta,\phi\right)+\sqrt{\frac{\ell+m+1}{2\ell+1}}\,_{q}Y_{\ell\,m+1}\left(\theta,\phi\right)
     \end{array}\right),\qquad\sqrt{\ell^{2}-q^{2}} \,,\\
   \chi_{q\ell m}^{\left(-\right)} &\equiv
    \frac{1}{\sqrt{2}} e^{i\alpha} R \left(\begin{array}{c}
       i\sqrt{\frac{\ell+m}{2\ell-1}}\,_{q}Y_{\ell-1\,m}\left(\theta,\phi\right)-\sqrt{\frac{\ell-m}{2\ell+1}}\,_{q}Y_{\ell m}\left(\theta,\phi\right)\\
       i\sqrt{\frac{\ell-m-1}{2\ell-1}}\,_{q}Y_{\ell-1\,m+1}\left(\theta,\phi\right)+\sqrt{\frac{\ell+m+1}{2\ell+1}}\,_{q}Y_{\ell\,m+1}\left(\theta,\phi\right)
     \end{array}\right),\qquad-\sqrt{\ell^{2}-q^{2}} \,.
 }
In \eqref{chipmDef} we included a phase $e^{i \alpha}$ that does not affect the eigenvalue equation nor the normalization of the eigenspinors.  If chosen appropriately, we can ensure that $\sigma_3 \chi^{(+)}_{q \ell m} =  \chi^{(-)}_{q \ell m}$, as mentioned in the main text;  this choice is
 \es{alphaFormula}{
  e^{2 i \alpha} =- \frac{q}{\ell} - i \sqrt{1 - \frac{q^2}{\ell^2}} \,.
 }

\section{Explicit expressions for $O(\epsilon^0)$ functional determinant}
\label{explicit}
We record the explicit values of the functions $\text{I}^{(q)}$ and $\text{II}^{(q)}$, as well as their $q=0$ values, used to compute the one loop $O(\epsilon^0)$ functional determinant in \eqref{FinalDet}. Note that we set $R_\HH=R_S=1$.
\es{Iasympex}{
\text{I}^{(q)}=&-\frac{N}{4 \pi }\left[ -{20q^2+39\ov 30}\log\mu  +4 \zeta '(-3,q+1)+2 \zeta
   (-3,q+1)+q \right.\\
&\left.  +   \left(q^2-1\right) \left(\left(q^2-1\right) \psi (q+1)-4
   \zeta '(-1,q+1)\right)\vphantom{\frac{1}{2}}\right.\\
   &\left.-\sum_{\ell=q+1}^\infty\left[2 \ell\left(\ell^2-q^2+1\right)
   \log \left(\frac{\ell^2+1-q^2}{\ell^2}\right)+\frac{2 \left(\ell^2+1\right) q^2}{\ell}-\frac{2
   \ell^2+q^4+1}{\ell}\right]\right]\,, \\
  \text{I}^{(0)}=&-\frac{N}{4\pi }\left[ -{39\ov 30}\log\mu  +4 \zeta '(-3,1)+4 \zeta '(-1,1)-\gamma +\frac{1}{60} \right.\\
&\left.-\sum_{\ell=1}^\infty\left[2 \ell \left(\ell^2+1\right)
   \log \left(\frac{\ell^2+1}{\ell^2}\right)-\frac{2 \ell^2+1}{\ell}\right]  \right]\,.
   }
   
 Term II is of the form
  \es{IIform}{
 \text{II}=&-\frac{16N}{45\pi }\,\log\mu -\frac{N}{\pi  }\left(\int_0^1 d\lambda \lambda \coth(\pi\lambda) \bullet+\int_1^\infty d\lambda \lambda \left(\coth(\pi\lambda)-1\right) \bullet\right)\text{ii}\,,
 }
 where the bullet stands for the integrand. We will therefore list only the integrands, denoted by lower case Roman numerals, which all give finite results when integrated in \eqref{IIform}:
 \es{IIasympex}{
 \text{ii}^{(q)}=& -2 \zeta '(-1,q+1)+\left(q^2-\lambda ^2\right) \psi (q+1)+q
   \log \lambda\\
   & + \sum_{\ell=q+1}^\infty \le[\ell \log\left({\ell^2+\lambda^2-q^2\ov \ell^2}\right)-{\lambda^2-q^2\ov \ell}\ri]\,,\\
 \text{ii}^{(0)}=&  \gamma  \lambda ^2-2 \zeta '(-1,1)+ \sum_{\ell=1}^\infty \le[\ell \log\left({\ell^2+\lambda^2\ov \ell^2}\right)-{\lambda^2\ov \ell}\ri]\,.
 }
In terms of these quantities, we express $ \mathfrak{f}(q)$ as
\es{fExplicitly}{
\mathfrak{f}(q)=\frac1N\left[\text{I}^{(q)}+\text{II}^{(q)}-\text{I}^{(0)}-\text{II}^{(0)}-\frac{Nq^2}{6\pi}\log\mu\right]+\frac{q^2\log(4\pi e^{-\gamma})}{12\pi}+O(\epsilon)\,.
}

\section{$O(\epsilon)$ functional determinant at large $q$}\label{app:largeq}
In the limit $q\to\infty$ we can evaluate the functional determinant in Section \ref{determinants} exactly. We will evaluate this determinant to order $O(q^{-1})$. In the following we set $R_S,R_\HH,\mu=1$, for simplicity.

Consider term I \eqref{ferm1}, except replacing $\ell\rightarrow n+q$ so that we can bring the $q$ dependence out of the sum, and letting the integration range be $(0,\infty)$:
  \es{largeQFermII}{
\text{I}^{(q)}&= \frac{N}{\pi}\dds\left(\sum_{n=1}^\infty (n+q)\int_0^\infty d\lambda \ \lambda \left(\lambda^2+n^2+2nq\right)^{-s} \right) \\
 &=\frac{N}{2\pi}\dds\sum_{n=1}^\infty \frac{n(n+q)(n+2q) \left(n^2+2qn\right)^{-s}}{s-1}\,,
 }
where we used that the $n=0$ mode doesn't contribute to term $\text{I}^{(q)}$.\footnote{In~\eqref{ferm1} the $\ell=q$ mode contributed, because the integration range was chosen to be $(1,\infty)$. }

Now that we have regulated the divergent integral, it is permissible to do a series expansion in large $q$:
 \es{largeQFermIIagain}{
\text{I}^{(q)}&=\frac{N}{2\pi}\dds\sum_{n=1}^\infty q^{-s}\left(q^2\, \frac{(2n)^{1-s}}{s-1}-q\,\frac{(2n)^{2-s}\,(s-3)}{4(s-1)}+{(2n)^{3-s}\,(s-4)\ov 32}+O(q^{-1})\right) \\
&=-\frac{N}{12\pi}\left[q^2\,\log\left(\frac{2q}{A^{12}}\right)+q\,18\zeta'(-2)-{\log(2q)\ov 20}+\frac{480\zeta'(-3)-1}{80}+O(q^{-1})\right] \,,
 }
where $A$ is the Glaisher constant.
 
Now consider term II~\eqref{fermSep}, again replacing $\ell\rightarrow n+q$ and letting the integration range be $(0,\infty)$. The integral is already convergent, so we can immediately expand for large $q$:
  \es{largeQFermI}{
\text{II}^{(q)}=& \frac{N}{\pi}\dds\left(\frac{q}{2}\int_0^\infty d\lambda \ \lambda^{1-2s} (\coth(\pi\lambda)-1)\right.\\
&\left.+\sum_{n=1}^\infty (n+q)\int_0^\infty d\lambda \ \lambda (\coth(\pi\lambda)-1)\left(\lambda^2+n^2+2nq\right)^{-s} \right) \\
 =&\frac{N\,q\le(\log A^{12}-1\ri)}{12\pi}\\
 &+\frac{N}{\pi}\dds\sum_{n=1}^\infty \int_0^\infty d\lambda \ \lambda (\coth(\pi\lambda)-1) q^{-s}\left[ q \,(2n)^{-s}-(2n)^{-(s+1)}(s(\lambda^2+n^2)-2n^2)+O(q^{-1}) \right]\\
 =&-\frac{N}{12\pi}\left[-{q\log \le(q\, A^{24} \ov \pi \ri)\ov 2}+q-\log(2q)\,{2\ov 15} +\frac{1}{40}\left(-5+2\gamma+40\log A\right)+O(q^{-1})\right] \,.
 }
Combining \eqref{largeQFermII} and \eqref{largeQFermI} we get:
\es{largqFinalFerm}{
\text{I}^{(q)}+\text{II}^{(q)}=&-\frac{N}{12\pi}\left[q^2\log\left(\frac{2q}{A^{12}}\right)-{q\log \le(q\, A^{24} \ov \pi \ri)\ov 2}+q\,\le(1-{9\zeta(3)\ov 2\pi^2}\ri)-{11\ov 60}\,\log(2q) \right.\\
&\left.+\le(-{11\ov 80}+{\gamma\ov 20}+\log A+6\zeta'(-3)\ri)+O(q^{-1})\right]\,.
}

 \section{Short distance behavior of the gauge field effective action}\label{app:ShortDist}

\subsection{Fermion propagator on $S^2 \times \HH^2$ when $q=0$} 

We find that writing the $S^2 \times \HH^2$ metric in the following form is most convenient:
 \es{Metric}{
  ds^2 = \frac{4 (dx_1^2 + dx_2^2)}{(1 + x_1^2 + x_2^2)^2} + \frac{dz^2 + dy^2}{z^2} \,.
 }
Here, $(x_1, x_2)$ are the stereographic coordinates on $S^2$ and $(z, y)$ are the Poincar\'e coordinates on $\HH^2$.  We use the frame
 \es{Frame}{
  e^1 = \frac{2 dx_1}{1 + x_1^2 + x_2^2} \,, \qquad
   e^2 = \frac{2 dx_2}{1 + x_1^2 + x_2^2}  \,, \qquad
   e^3 = \frac{dz}{z} \,, \qquad e^4 = \frac{dy}{z} 
 }
and the same gamma matrices as in~\eqref{gammas}.

The metric on the sphere can be put in the standard form $ds^2 = d\theta^2 + \sin^2 \theta\, d\phi^2$ if we define 
 \es{xToAngles}{
   x_1 = \tan \frac{\theta}{2} \cos \phi \,, \qquad
    x_2 = \tan \frac{\theta}{2} \sin \phi \,.
 }

The propagator between $(x_1, x_2, z, y)$ and $(0, 0, 1, 0)$ is
 \es{PropagFree}{
  G_0 &= \frac{z^{3/2} (1 + x_1^2 + x_2^2)^{3/2}}{2\pi^2 \left[ (x_1^2 + x_2^2) (y^2 + (1 + z)^2) + y^2 + (z-1)^2 \right]^2 } \\
   &\times \left[\sigma_3 \otimes ((z-1) \sigma_1 + y \sigma_2)
    + (x_1 \sigma_1 + x_2 \sigma_2) \otimes ((z+1) \mathbf{1} - y i \sigma_3 ) \right] \,.
 }
One can check that $i \slashed{\nabla} G = 0$ at separated points.   It is normalized so that it gives a delta-function of unit strength.

\subsection{Fermion propagator when $q \neq 0$}

When $q \neq 0$, one can find a series expansion for the Green's function.  To simplify the expressions, let's define
 \es{XZDef}{
  X^2 = 4 (x_1^2 + x_2^2) \,, \qquad
   Z^2 = \frac{y^2 + (z-1)^2}{z} \,,
 }
as well as
 \es{SDef}{
  S^2 = X^2 + Z^2 + X^2 Z^2 / 4 \,,
 } 
which is the quantity that appears in the denominator of the $q=0$ Green's function \eqref{PropagFree}.   Then expanding at small $S$, we have
 \es{GExpansion}{
  G_q &= \frac{ (1 + X^2/4)^{3/2}}{2 \pi^2 \sqrt{z}  } \times
   \biggl[
    \left( \sigma_3 \frac{1}{S^4} + {\bf 1} \frac{q}{4 S^2}  
     - \sigma_3 \frac{q^2}{48} \left(-\frac{Z^2}{S^2} + \log S^2 + c \right) + \cdots
      \right) \otimes ((z-1) \sigma_1 + y \sigma_2)  \biggr]
     \\
   &+ (x_1 \sigma_1 + x_2 \sigma_2) \otimes ((z+1) \mathbf{1} - y i \sigma_3 )  
    \left[\frac{1}{S^4} + \frac{q^2}{48} \left( - \frac{X^2}{S^2} +  \log S^2 + c \right) + \cdots \right] \\
   &+G_\text{analytic} \,, 
 }
where $c$ is an arbitrary constant and $G_\text{analytic}$ is an analytic piece.  (The constant $c$ multiplies analytic terms also.) This expression was obtained from requiring $G_q$ to satisfy $i \slashed{\nabla} G = 0$ at separated points.

\subsection{Contribution to the gauge field effective action}

Let's construct
 \es{KDef}{
  K^q_{ij}(x, x')  = -\langle J_i(x) J_j(x') \rangle = \tr[ \gamma_i G_q(x, x') \gamma_j G_q^\dagger(x, x')] \,.
 }
(Note that if $\langle \psi(x) \psi^\dagger(x') \rangle = G(x, x')$, then $\langle \psi^\dagger(x) \psi(x') \rangle = -G(x, x')^\dagger$ because complex conjugation interchanges the order of the fermions.)  We want the difference $K^q_{ij} - K^0_{ij}$ to leading order in $S$.  For that, define the coordinates $y^1 = 2 x^1$, $y^2 = 2x^2$, $y^3 = z-1$, $y^4 = y$ and the operators
 \es{Proj}{
  \Pi_{ij} &= \delta_{ij} \sum_{k=1}^4 \partial_k \partial^k - \partial_i \partial_j  \,, \\
   \Pi^S_{ij} &= \begin{cases}
    \delta_{ij} \sum_{k=1}^2 \partial_k \partial^k - \partial_i \partial_j & \text{if $i, j \leq 2$} \\
    0 & \text{otherwise}
   \end{cases} \,, \\
     \Pi^H_{ij} &= \begin{cases}
    \delta_{ij} \sum_{k=3}^4 \partial_k \partial^k - \partial_i \partial_j & \text{if $i, j \geq 3$} \\
    0 & \text{otherwise}
   \end{cases} 
 }
Then
 \es{KDiff}{
  K^q_{ij} - K^0_{ij}
   &= \frac{1}{4\pi^4} \Pi_{ij} \left(-\frac{q^2 Z^2}{24 (X^2 + Z^2)} \right)
    +\frac{1}{4 \pi^4} \Pi^S_{ij} \left(\frac{q^2 (\log^2 (X^2 + Z^2) + 2 c \log (X^2 + Z^2) }{24}  \right) \\
     &+\frac{1}{4\pi^4} \Pi^H_{ij} \left(-\frac{q^2 \left[ \log^2 (X^2 + Z^2) + (2c - 3)  \log (X^2 + Z^2) \right]}{24} \right) \,.
 }

We're only interested in $\delta^{ij} K_{ij}$, so 
 \es{KDiffDelta}{
  K^q_{ii} - K^0_{ii}
   &=  -\frac{3}{4 \pi^4} \nabla^2 \left(\frac{q^2 Z^2}{24 (X^2 + Z^2)} \right)
    +\frac{1}{4 \pi^4} \nabla_S^2 \left(\frac{q^2 (\log^2 (X^2 + Z^2) + 2 c \log (X^2 + Z^2) }{24}  \right) \\
     &+\frac{1}{4 \pi^4} \nabla_H^2 \left(-\frac{q^2 \left[ \log^2 (X^2 + Z^2) + (2c - 3)  \log (X^2 + Z^2) \right]}{24} \right) \,.
 }
 At small $X$ and $Z$, one can ignore curvature effects, so \eqref{KDiffDelta} becomes an expression in flat space.
  This expression can be transformed to Fourier space with the help of the formulas
 \es{FT}{
  \int d^4 x \frac{e^{i p \cdot x}}{\abs{x}^2} &= \frac{4 \pi^2}{\abs{p}^2} \,, \\
  \int d^4 x e^{i p \cdot x} \log \abs{x}^2 &= -\frac{16 \pi^2}{\abs{p}^4} \,, \\
  \int d^4 x e^{i p \cdot x} \log^2 \abs{x}^2 &= 32 \pi^2 \frac{-1 + 2\gamma + \log (\abs{p}^2/4) }{\abs{p}^4} \,.
 }
With the notation $p_S^2 = p_1^2 + p_2^2$ and $p_H^2 = p_3^2 + p_4^2$, we obtain
 \es{KMom}{
  K^q_{ii}(p) - K^0_{ii}(p) 
   &= \frac{1}{4\pi^4}p^2 \left(\frac{q^2 (p_S^2 - p_H^2) }{2 p^6} \right)
    -\frac{1}{4\pi^4}(p_S^2 -p_H^2) \frac{q^2}{24} \left(32 \pi^2 \frac{-1 - c + 2\gamma + \log (\abs{p}^2/4) }{\abs{p}^4} \right) \\
     & + q^2  \frac{2 p_H^2 }{4 \pi^2 \abs{p}^4} \,.
 }

\bibliographystyle{ssg}
\bibliography{MonopoleEpsilon}

\begingroup\raggedright\begin{thebibliography}{10}

\bibitem{Polyakov:1975rs}
A.~M. Polyakov, ``{Compact gauge fields and the infrared catastrophe},'' {\em
  Phys.Lett.} {\bf B59} (1975) 82--84.

\bibitem{Wen:1993zza}
X.-G. Wen and Y.-S. Wu, ``{Transitions between the quantum Hall states and
  insulators induced by periodic potentials},'' {\em Phys.Rev.Lett.} {\bf 70}
  (1993) 1501--1504.

\bibitem{Chen:1993cd}
W.~Chen, M.~P. Fisher, and Y.-S. Wu, ``{Mott transition in an anyon gas},''
  {\em Phys.Rev.} {\bf B48} (1993) 13749--13761,
  \href{http://xxx.lanl.gov/abs/cond-mat/9301037}{{\tt cond-mat/9301037}}.

\bibitem{Sachdev97}
S.~Sachdev, ``{Non-zero temperature transport near fractional quantum Hall
  critical points},'' {\em Phys.Rev.} {\bf B57} (1998) 7157.

\bibitem{Rantner01}
W.~Rantner and X.-G. Wen, ``{Electron spectral function and algebraic spin
  liquid for the normal state of underdoped high $T_c$ superconductors},'' {\em
  Phys.Rev.Lett.} {\bf 86} (2001) 3871.

\bibitem{Rantner:2002zz}
W.~Rantner and X.-G. Wen, ``{Spin correlations in the algebraic spin liquid:
  Implications for high-$T_c$ superconductors},'' {\em Phys.Rev.} {\bf B66}
  (2002) 144501.

\bibitem{Motrunich:2003fz}
O.~I. Motrunich and A.~Vishwanath, ``{Emergent photons and new transitions in
  the O(3) sigma model with hedgehog suppression},'' {\em Phys.Rev.} {\bf B70}
  (2004) 075104, \href{http://xxx.lanl.gov/abs/cond-mat/0311222}{{\tt
  cond-mat/0311222}}.

\bibitem{SVBSF}
T.~{Senthil}, A.~{Vishwanath}, L.~{Balents}, S.~{Sachdev}, and M.~P.~A.
  {Fisher}, ``{Deconfined Quantum Critical Points},'' {\em Science} {\bf 303}
  (Mar., 2004) 1490--1494,
  \href{http://xxx.lanl.gov/abs/arXiv:cond-mat/0311326}{{\tt
  arXiv:cond-mat/0311326}}.

\bibitem{SBSVF}
T.~{Senthil}, L.~{Balents}, S.~{Sachdev}, A.~{Vishwanath}, and M.~P.~A.
  {Fisher}, ``{Quantum criticality beyond the Landau-Ginzburg-Wilson
  paradigm},'' {\em Phys.Rev.} {\bf 70} (Oct., 2004) 144407,
  \href{http://xxx.lanl.gov/abs/arXiv:cond-mat/0312617}{{\tt
  arXiv:cond-mat/0312617}}.

\bibitem{Hermele}
M.~{Hermele}, T.~{Senthil}, M.~P.~A. {Fisher}, P.~A. {Lee}, N.~{Nagaosa}, and
  X.-G. {Wen}, ``{Stability of U(1) spin liquids in two dimensions},'' {\em
  Phys.Rev.} {\bf B70} (2004) 214437,
  \href{http://xxx.lanl.gov/abs/arXiv:cond-mat/0404751}{{\tt
  arXiv:cond-mat/0404751}}.

\bibitem{Hermele05}
M.~Hermele, T.~Senthil, and M.~P. Fisher, ``{Algebraic spin liquid as the
  mother of many competing orders},'' {\em Phys.Rev.} {\bf B72} (2005) 104404.

\bibitem{Ran06}
Y.~{Ran} and X.-G. Wen, ``{Continuous quantum phase transitions beyond Landau's
  paradigm in a large-$N$ spin model},''
  \href{http://xxx.lanl.gov/abs/cond-mat/0609620}{{\tt cond-mat/0609620}}.

\bibitem{Kaul08}
R.~K. Kaul, Y.~B. Kim, S.~Sachdev, and T.~Senthil, ``{Algebraic charge
  liquids},'' {\em Nature Physics} {\bf 4} (2008) 28--31.

\bibitem{Kaul:2008xw}
R.~K. Kaul and S.~Sachdev, ``{Quantum criticality of U(1) gauge theories with
  fermionic and bosonic matter in two spatial dimensions},'' {\em Phys.Rev.}
  {\bf B77} (2008) 155105, \href{http://xxx.lanl.gov/abs/0801.0723}{{\tt
  0801.0723}}.

\bibitem{Sachdev:2010uz}
S.~Sachdev, ``{The landscape of the Hubbard model},''
  \href{http://xxx.lanl.gov/abs/1012.0299}{{\tt 1012.0299}}.

\bibitem{Borokhov:2002ib}
V.~Borokhov, A.~Kapustin, and X.-k. Wu, ``{Topological disorder operators in
  three-dimensional conformal field theory},'' {\em JHEP} {\bf 0211} (2002)
  049, \href{http://xxx.lanl.gov/abs/hep-th/0206054}{{\tt hep-th/0206054}}.

\bibitem{Murthy:1989ps}
G.~Murthy and S.~Sachdev, ``{Action of hedgehog instantons in the disordered
  phase of the $(2+1)$-dimensional $\CP^{N-1}$ model},'' {\em Nucl.Phys.} {\bf
  B344} (1990) 557--595.

\bibitem{Pufu:2013vpa}
S.~S. Pufu, ``{Anomalous dimensions of monopole operators in three-dimensional
  quantum electrodynamics},'' {\em Phys.Rev.} {\bf D89} (2014), no.~6 065016,
  \href{http://xxx.lanl.gov/abs/1303.6125}{{\tt 1303.6125}}.

\bibitem{Dyer:2013fja}
E.~Dyer, M.~Mezei, and S.~S. Pufu, ``{Monopole Taxonomy in Three-Dimensional
  Conformal Field Theories},'' \href{http://xxx.lanl.gov/abs/1309.1160}{{\tt
  1309.1160}}.

\bibitem{Benini:2009qs}
F.~Benini, C.~Closset, and S.~Cremonesi, ``{Chiral flavors and M2-branes at
  toric CY4 singularities},'' {\em JHEP} {\bf 1002} (2010) 036,
  \href{http://xxx.lanl.gov/abs/0911.4127}{{\tt 0911.4127}}.

\bibitem{Benini:2011cma}
F.~Benini, C.~Closset, and S.~Cremonesi, ``{Quantum moduli space of
  Chern-Simons quivers, wrapped D6-branes and AdS4/CFT3},'' {\em JHEP} {\bf
  1109} (2011) 005, \href{http://xxx.lanl.gov/abs/1105.2299}{{\tt 1105.2299}}.

\bibitem{2013PhRvL.111m7202B}
M.~S. {Block}, R.~G. {Melko}, and R.~K. {Kaul}, ``{Fate of CP$^{N-1}$ Fixed
  Points with q Monopoles},'' {\em Physical Review Letters} {\bf 111} (Sept.,
  2013) 137202, \href{http://xxx.lanl.gov/abs/1307.0519}{{\tt 1307.0519}}.

\bibitem{2015arXiv150205128K}
R.~K. {Kaul} and M.~{Block}, ``{Numerical studies of various Neel-VBS
  transitions in SU($N$) antiferromagnets},''
  \href{http://xxx.lanl.gov/abs/1502.05128}{{\tt 1502.05128}}.

\bibitem{Rattazzi:2008pe}
R.~Rattazzi, V.~S. Rychkov, E.~Tonni, and A.~Vichi, ``{Bounding scalar operator
  dimensions in 4D CFT},'' {\em JHEP} {\bf 0812} (2008) 031,
  \href{http://xxx.lanl.gov/abs/0807.0004}{{\tt 0807.0004}}.

\bibitem{Chester:2016wrc} 
  S.~M.~Chester and S.~S.~Pufu, ``{Towards bootstrapping QED$_{3}$},''
  {\em JHEP} {\bf 1608} (2016) 019,
 \href{http://xxx.lanl.gov/abs/1601.03476}{{\tt 1601.03476}}.

\bibitem{Hellerman:2015nra}
S.~Hellerman, D.~Orlando, S.~Reffert, and M.~Watanabe, ``{On the CFT Operator
  Spectrum at Large Global Charge},''
  \href{http://xxx.lanl.gov/abs/1505.01537}{{\tt 1505.01537}}.

\bibitem{Wilson:1973jj}
K.~G. Wilson and J.~B. Kogut, ``{The Renormalization group and the epsilon
  expansion},'' {\em Phys. Rept.} {\bf 12} (1974) 75--200.

\bibitem{Giombi:2015haa}
S.~Giombi, I.~R. Klebanov, and G.~Tarnopolsky, ``{Conformal QED$_d$,
  $F$-Theorem and the $\epsilon$ Expansion},''
  \href{http://xxx.lanl.gov/abs/1508.06354}{{\tt 1508.06354}}.

\bibitem{Fei:2015oha}
L.~Fei, S.~Giombi, I.~R. Klebanov, and G.~Tarnopolsky, ``{Generalized
  $F$-Theorem and the $\epsilon$ Expansion},''
  \href{http://xxx.lanl.gov/abs/1507.01960}{{\tt 1507.01960}}.

\bibitem{Fei:2014yja}
L.~Fei, S.~Giombi, and I.~R. Klebanov, ``{Critical $O(N)$ models in
  $6-\epsilon$ dimensions},'' {\em Phys. Rev.} {\bf D90} (2014), no.~2 025018,
  \href{http://xxx.lanl.gov/abs/1404.1094}{{\tt 1404.1094}}.

\bibitem{Giombi:2014xxa}
S.~Giombi and I.~R. Klebanov, ``{Interpolating between $a$ and $F$},'' {\em
  JHEP} {\bf 03} (2015) 117, \href{http://xxx.lanl.gov/abs/1409.1937}{{\tt
  1409.1937}}.

\bibitem{Fei:2014xta}
L.~Fei, S.~Giombi, I.~R. Klebanov, and G.~Tarnopolsky, ``{Three loop analysis
  of the critical O(N) models in 6-ε dimensions},'' {\em Phys. Rev.} {\bf D91}
  (2015), no.~4 045011, \href{http://xxx.lanl.gov/abs/1411.1099}{{\tt
  1411.1099}}.

\bibitem{Fei:2015kta}
L.~Fei, S.~Giombi, I.~R. Klebanov, and G.~Tarnopolsky, ``{Critical Sp(N )
  models in 6 − ϵ dimensions and higher spin dS/CFT},'' {\em JHEP} {\bf 09}
  (2015) 076, \href{http://xxx.lanl.gov/abs/1502.07271}{{\tt 1502.07271}}.

\bibitem{Gaiotto:2013nva}
D.~Gaiotto, D.~Mazac, and M.~F. Paulos, ``{Bootstrapping the 3d Ising twist
  defect},'' {\em JHEP} {\bf 03} (2014) 100,
  \href{http://xxx.lanl.gov/abs/1310.5078}{{\tt 1310.5078}}.

\bibitem{Gaiotto:2014kfa}
D.~Gaiotto, A.~Kapustin, N.~Seiberg, and B.~Willett, ``{Generalized Global
  Symmetries},'' {\em JHEP} {\bf 02} (2015) 172,
  \href{http://xxx.lanl.gov/abs/1412.5148}{{\tt 1412.5148}}.

\bibitem{Kapustin:2005py}
A.~Kapustin, ``{Wilson-'t Hooft operators in four-dimensional gauge theories
  and S-duality},'' {\em Phys.Rev.} {\bf D74} (2006) 025005,
  \href{http://xxx.lanl.gov/abs/hep-th/0501015}{{\tt hep-th/0501015}}.

\bibitem{Moshe:2003xn}
M.~Moshe and J.~Zinn-Justin, ``{Quantum field theory in the large N limit: A
  Review},'' {\em Phys. Rept.} {\bf 385} (2003) 69--228,
  \href{http://xxx.lanl.gov/abs/hep-th/0306133}{{\tt hep-th/0306133}}.

\bibitem{Redlich:1983dv}
A.~Redlich, ``{Parity Violation and Gauge Noninvariance of the Effective Gauge
  Field Action in Three-Dimensions},'' {\em Phys.Rev.} {\bf D29} (1984)
  2366--2374.

\bibitem{Redlich:1983kn}
A.~Redlich, ``{Gauge Noninvariance and Parity Violation of Three-Dimensional
  Fermions},'' {\em Phys.Rev.Lett.} {\bf 52} (1984) 18.

\bibitem{Niemi:1983rq}
A.~Niemi and G.~Semenoff, ``{Axial Anomaly Induced Fermion Fractionization and
  Effective Gauge Theory Actions in Odd Dimensional Space-Times},'' {\em
  Phys.Rev.Lett.} {\bf 51} (1983) 2077.

\bibitem{Dyer:2015zha}
E.~Dyer, M.~Mezei, S.~S. Pufu, and S.~Sachdev, ``{Scaling dimensions of
  monopole operators in the $ \mathbb{C}{\mathrm{\mathbb{P}}}^{N_b-1} $ theory
  in 2 $+$ 1 dimensions},'' {\em JHEP} {\bf 06} (2015) 037,
  \href{http://xxx.lanl.gov/abs/1504.00368}{{\tt 1504.00368}}.

\bibitem{Komargodski:2011vj}
Z.~Komargodski and A.~Schwimmer, ``{On Renormalization Group Flows in Four
  Dimensions},'' {\em JHEP} {\bf 12} (2011) 099,
  \href{http://xxx.lanl.gov/abs/1107.3987}{{\tt 1107.3987}}.

\bibitem{Cognola:2003zt}
G.~Cognola and S.~Zerbini, ``{Effective action for scalar fields and
  generalized zeta function regularization},'' {\em Phys. Rev.} {\bf D69}
  (2004) 024004, \href{http://xxx.lanl.gov/abs/hep-th/0309221}{{\tt
  hep-th/0309221}}.

\bibitem{Bytsenko:1994bc}
A.~A. Bytsenko, G.~Cognola, L.~Vanzo, and S.~Zerbini, ``{Quantum fields and
  extended objects in space-times with constant curvature spatial section},''
  {\em Phys. Rept.} {\bf 266} (1996) 1--126,
  \href{http://xxx.lanl.gov/abs/hep-th/9505061}{{\tt hep-th/9505061}}.

\bibitem{Gorishnii:1991hw}
S.~G. Gorishnii, A.~L. Kataev, and S.~A. Larin, ``{The three loop QED
  contributions to the photon vacuum polarization function in the MS scheme and
  the four loop corrections to the QED beta function in the on-shell scheme},''
  {\em Phys. Lett.} {\bf B273} (1991) 141--144. [Erratum: Phys.
  Lett.B341,448(1995)].

\bibitem{Pufu:2013eda}
S.~S. Pufu and S.~Sachdev, ``{Monopoles in $2 + 1$-dimensional conformal field
  theories with global U(1) symmetry},'' {\em JHEP} {\bf 1309} (2013) 127,
  \href{http://xxx.lanl.gov/abs/1303.3006}{{\tt 1303.3006}}.

\bibitem{Hawking:1976ja}
S.~W. Hawking, ``{Zeta Function Regularization of Path Integrals in Curved
  Space-Time},'' {\em Commun. Math. Phys.} {\bf 55} (1977) 133.

\bibitem{DiPietro:2015taa}
L.~Di~Pietro, Z.~Komargodski, I.~Shamir, and E.~Stamou, ``{Quantum
  Electrodynamics in d=3 from the epsilon-expansion},''
  \href{http://xxx.lanl.gov/abs/1508.06278}{{\tt 1508.06278}}.

\bibitem{Camporesi:1995fb}
R.~Camporesi and A.~Higuchi, ``{On the Eigen functions of the Dirac operator on
  spheres and real hyperbolic spaces},'' {\em J. Geom. Phys.} {\bf 20} (1996)
  1--18, \href{http://xxx.lanl.gov/abs/gr-qc/9505009}{{\tt gr-qc/9505009}}.

\end{thebibliography}\endgroup

\end{document}